\documentclass[a4paper,11pt]{article}
\pdfoutput=1 

\usepackage{jcappub} 

\usepackage[T1]{fontenc} 

\usepackage{graphicx}
\usepackage[utf8]{inputenc}
\usepackage[english]{babel}   
\usepackage{amssymb}
\usepackage{amsmath}
\usepackage{slashed}
\usepackage{color}
\usepackage{mathtools}
\usepackage{hyperref}
\usepackage{floatrow}
\usepackage{makecell}
\usepackage{subfigure}
\usepackage{multirow}
\usepackage{booktabs}
\usepackage{enumitem}
\usepackage{placeins}
\usepackage{arydshln}
\usepackage{aas_macros}
\newcommand{\sorthelp}[1]{}
\usepackage{lineno}

\newcolumntype{P}[1]{>{\centering\arraybackslash}p{#1}}

\title{\boldmath Requirements on bandpass resolution and measurement precision for LiteBIRD}

\author[1]{S.\,Giardiello,\note{Corresponding author.}}
\author[2,3]{A.\,Carones,}
\author[4,5]{T.\,Ghigna,}
\author[6,7,8]{L.\,Pagano,}
\author[9,10]{F.\,Piacentini,}
\author[11]{L.\,Montier,}
\author[12,13]{R.\,Takaku,}
\author[1]{E.\,Calabrese,}
\author[14]{D.\,Adak,}
\author[15]{E.\,Allys,}
\author[16]{A.\,Anand,}
\author[11]{J.\,Aumont,}
\author[6,7,17]{M.\,Ballardini,}
\author[11]{A.\,J.\,Banday,}
\author[18]{R.\,B.\,Barreiro,}
\author[19,20,21]{N.\,Bartolo,}
\author[22]{S.\,Basak,}
\author[23,24]{M.\,Bersanelli,}
\author[8]{A.\,Besnard,}
\author[6,7]{M.\,Bortolami,}
\author[6]{T.\,Brinckmann,}
\author[18]{F.\,J.\,Casas,}
\author[25,26,27,28]{K.\,Cheung,}
\author[29]{M.\,Citran,}
\author[30]{L.\,Clermont,}
\author[9,10]{F.\,Columbro,}
\author[9,10]{A.\,Coppolecchia,}
\author[17]{F.\,Cuttaia,}
\author[9,10]{P.\,de\,Bernardis,}
\author[31]{E.\,de\,la\,Hoz,}
\author[32]{M.\,De\,Lucia,}
\author[33]{S.\,Della\,Torre,}
\author[32]{E.\,Di\,Giorgi,}
\author[34]{P.\,Diego-Palazuelos,}
\author[35]{U.\,Fuskeland,}
\author[6,16]{G.\,Galloni,}
\author[35]{M.\,Galloway,}
\author[7]{M.\,Gerbino,}
\author[36,33]{M.\,Gervasi,}
\author[14,37]{R.\,T.\,Génova-Santos,}
\author[18]{C.\,Gimeno-Amo,}
\author[17,38]{A.\,Gruppuso,}
\author[39,13,5,40]{M.\,Hazumi,}
\author[41]{S.\,Henrot-Versillé,}
\author[42,41]{L.\,T.\,Hergt,}
\author[5]{B.\,Jost,}
\author[39]{K.\,Kohri,}
\author[9,10]{L.\,Lamagna,}
\author[5]{C.\,Leloup,}
\author[15]{F.\,Levrier,}
\author[43]{A.\,I.\,Lonappan,}
\author[44,45]{M.\,López-Caniego,}
\author[46]{G.\,Luzzi,}
\author[47]{J.\,Macias-Perez,}
\author[46,9,16]{V.\,Maranchery,}
\author[18]{E.\,Martínez-González,}
\author[9,10]{S.\,Masi,}
\author[19,20,21,48]{S.\,Matarrese,}
\author[5]{T.\,Matsumura,}
\author[9]{S.\,Micheli,}
\author[16,49]{M.\,Migliaccio,}
\author[5]{M.\,Monelli,}
\author[17]{G.\,Morgante,}
\author[15,11]{L.\,Mousset,}
\author[13]{R.\,Nagata,}
\author[9]{A.\,Novelli,}
\author[1]{F.\,Noviello,}
\author[39]{I.\,Obata,}
\author[9]{A.\,Occhiuzzi,}
\author[9,10]{A.\,Paiella,}
\author[17,38]{D.\,Paoletti,}
\author[5,18]{G.\,Pascual-Cisneros,}
\author[29]{G.\,Patanchon,}
\author[32]{M.\,Pinchera,}
\author[46]{G.\,Polenta,}
\author[50]{L.\,Porcelli,}
\author[51,52,53]{G.\,Puglisi,}
\author[6]{N.\,Raffuzzi,}
\author[18]{M.\,Remazeilles,}
\author[54,29]{A.\,Rizzieri,}
\author[18,55]{M.\,Ruiz-Granda,}
\author[56,57]{J.\,Sanghavi,}
\author[8]{V.\,Sauvage,}
\author[58]{G.\,Savini,}
\author[59]{M.\,Shiraishi,}
\author[60,32]{G.\,Signorelli,}
\author[35]{R.\,M.\,Sullivan,}
\author[61]{Y.\,Takase,}
\author[17]{L.\,Terenzi,}
\author[23,24]{M.\,Tomasi,}
\author[41]{M.\,Tristram,}
\author[2]{L.\,Vacher,}
\author[41]{B.\,van\,Tent,}
\author[18]{P.\,Vielva,}
\author[35]{I.\,K.\,Wehus,}
\author[54,41]{G.\,Weymann-Despres,}
\author[62]{E.\,J.\,Wollack,}
\author[4]{and Y.\,Zhou}
\author[ ]{\\LiteBIRD Collaboration.}
\affiliation[1]{School of Physics and Astronomy, Cardiff University, Cardiff CF24 3AA, UK}
\affiliation[2]{International School for Advanced Studies (SISSA), Via Bonomea 265, 34136, Trieste, Italy}
\affiliation[3]{INFN Sezione di Trieste, via Valerio 2, 34127 Trieste, Italy}
\affiliation[4]{International Center for Quantum-field Measurement Systems for Studies of the Universe and Particles (QUP), High Energy Accelerator Research Organization (KEK), Tsukuba, Ibaraki 305-0801, Japan}
\affiliation[5]{Kavli Institute for the Physics and Mathematics of the Universe (Kavli IPMU, WPI), UTIAS, The University of Tokyo, Kashiwa, Chiba 277-8583, Japan}
\affiliation[6]{Dipartimento di Fisica e Scienze della Terra, Università di Ferrara, Via Saragat 1, 44122 Ferrara, Italy}
\affiliation[7]{INFN Sezione di Ferrara, Via Saragat 1, 44122 Ferrara, Italy}
\affiliation[8]{Université Paris-Saclay, CNRS, Institut d’Astrophysique Spatiale, 91405, Orsay, France}
\affiliation[9]{Dipartimento di Fisica, Università La Sapienza, P. le A. Moro 2, Roma, Italy}
\affiliation[10]{INFN Sezione di Roma, P.le A. Moro 2, 00185 Roma, Italy}
\affiliation[11]{IRAP, Université de Toulouse, CNRS, CNES, UPS, Toulouse, France}
\affiliation[12]{The University of Tokyo, Department of Physics, Tokyo 113-0033, Japan}
\affiliation[13]{Japan Aerospace Exploration Agency (JAXA), Institute of Space and Astronautical Science (ISAS), Sagamihara, Kanagawa 252-5210, Japan}
\affiliation[14]{Instituto de Astrofísica de Canarias, E-38200 La Laguna, Tenerife, Canary Islands, Spain}
\affiliation[15]{Laboratoire de Physique de l’École Normale Supérieure, ENS, Université PSL, CNRS, Sorbonne Université, Université de Paris, 75005 Paris, France}
\affiliation[16]{Dipartimento di Fisica, Università di Roma Tor Vergata, Via della Ricerca Scientifica, 1, 00133, Roma, Italy}
\affiliation[17]{INAF - OAS Bologna, via Piero Gobetti, 93/3, 40129 Bologna, Italy}
\affiliation[18]{Instituto de Fisica de Cantabria (IFCA, CSIC-UC), Avenida los Castros SN, 39005, Santander, Spain}
\affiliation[19]{Dipartimento di Fisica e Astronomia “G. Galilei”, Università degli Studi di Padova, via Marzolo 8, I-35131 Padova, Italy}
\affiliation[20]{INFN Sezione di Padova, via Marzolo 8, I-35131, Padova, Italy}
\affiliation[21]{INAF, Osservatorio Astronomico di Padova, Vicolo dell’Osservatorio 5, I-35122, Padova, Italy}
\affiliation[22]{School of Physics, Indian Institute of Science Education and Research Thiruvananthapuram, Maruthamala PO, Vithura, Thiruvananthapuram 695551, Kerala, India}
\affiliation[23]{Dipartimento di Fisica, Università degli Studi di Milano, Via Celoria 16 - 20133, Milano, Italy}
\affiliation[24]{INFN Sezione di Milano, Via Celoria 16 - 20133, Milano, Italy}
\affiliation[25]{Jodrell Bank Centre for Astrophysics, Alan Turing Building, Department of Physics and Astronomy, School of Natural Sciences, The University of Manchester, Oxford Road, Manchester M13 9PL, UK}
\affiliation[26]{University of California, Berkeley, Department of Physics, Berkeley, CA 94720, USA}
\affiliation[27]{University of California, Berkeley, Space Sciences Laboratory,  Berkeley, CA 94720, USA}
\affiliation[28]{Lawrence Berkeley National Laboratory (LBNL), Computational Cosmology Center, Berkeley, CA 94720, USA}
\affiliation[29]{Université Paris Cité, CNRS, Astroparticule et Cosmologie, F-75013 Paris, France}
\affiliation[30]{Centre Spatial de Liège, Université de Liège, Avenue du Pré-Aily, 4031 Angleur, Belgium}
\affiliation[31]{CNRS-UCB International Research Laboratory, Centre Pierre Binétruy, UMI2007, Berkeley, CA 94720, USA}
\affiliation[32]{INFN Sezione di Pisa, Largo Bruno Pontecorvo 3, 56127 Pisa, Italy}
\affiliation[33]{INFN Sezione Milano Bicocca, Piazza della Scienza, 3, 20126 Milano, Italy}
\affiliation[34]{Max Planck Institute for Astrophysics, Karl-Schwarzschild-Str. 1, D-85748 Garching, Germany}
\affiliation[35]{Institute of Theoretical Astrophysics, University of Oslo, Blindern, Oslo, Norway}
\affiliation[36]{University of Milano Bicocca, Physics Department, p.zza della Scienza, 3, 20126 Milan, Italy}
\affiliation[37]{Departamento de Astrofísica, Universidad de La Laguna (ULL), E-38206, La Laguna, Tenerife, Spain}
\affiliation[38]{INFN Sezione di Bologna, Viale C. Berti Pichat, 6/2 – 40127 Bologna, Italy}
\affiliation[39]{Institute of Particle and Nuclear Studies (IPNS), High Energy Accelerator Research Organization (KEK), Tsukuba, Ibaraki 305-0801, Japan}
\affiliation[40]{The Graduate University for Advanced Studies (SOKENDAI), Miura District, Kanagawa 240-0115, Hayama, Japan}
\affiliation[41]{Université Paris-Saclay, CNRS/IN2P3, IJCLab, 91405 Orsay, France}
\affiliation[42]{Department of Physics and Astronomy, University of British Columbia, 6224 Agricultural Road, Vancouver, BC V6T1Z1, Canada}
\affiliation[43]{University of California, San Diego, Department of Physics, San Diego, CA 92093-0424, USA}
\affiliation[44]{Aurora Technology for the European Space Agency, Camino bajo del Castillo, s/n, Urbanización Villafranca del Castillo, Villanueva de la Cañada, Madrid, Spain}
\affiliation[45]{Universidad Europea de Madrid, 28670, Madrid, Spain}
\affiliation[46]{Space Science Data Center, Italian Space Agency, via del Politecnico, 00133, Roma, Italy}
\affiliation[47]{Université Grenoble Alpes, CNRS, LPSC-IN2P3, 53, avenue des Martyrs, 38000 Grenoble, France}
\affiliation[48]{Gran Sasso Science Institute (GSSI), Viale F. Crispi 7, I-67100, L’Aquila, Italy}
\affiliation[49]{INFN Sezione di Roma2, Università di Roma Tor Vergata, via della Ricerca Scientifica, 1, 00133 Roma, Italy}
\affiliation[50]{Istituto Nazionale di Fisica Nucleare–Laboratori Nazionali di Frascati (INFN–LNF), Via E. Fermi 40, 00044, Frascati, Italy}
\affiliation[51]{Dipartimento di Fisica e Astronomia, Universitá degli Studi di Catania, Via S. Sofia,64, 95123, Catania, Italy}
\affiliation[52]{INAF, Osservatorio Astrofisico di Catania, via S.Sofia 78, I-95123 Catania, Italy}
\affiliation[53]{INFN, Sezione di Catania, via S.Sofia 64, I-95123, Catania, Italy}
\affiliation[54]{Department of Physics, University of Oxford, Denys Wilkinson Building, Keble Road, Oxford OX1 3RH, UK}
\affiliation[55]{Dpto. de Física Moderna, Universidad de Cantabria, Avda. los Castros s/n, E-39005 Santander, Spain}
\affiliation[56]{Universitäts-Sternwarte, Fakultät für Physik, Ludwig-Maximilians Universität München, Scheinerstr.1, 81679 München, Germany}
\affiliation[57]{GRAPPA, Institute for Theoretical Physics Amsterdam, University of Amsterdam, Science Park 904, 1098 XH Amsterdam, The Netherlands}
\affiliation[58]{Physics and Astronomy Dept., University College London (UCL), UK}
\affiliation[59]{Suwa University of Science, Chino, Nagano 391-0292, Japan}
\affiliation[60]{Dipartimento di Fisica, Università di Pisa, Largo B. Pontecorvo 3, 56127 Pisa, Italy}
\affiliation[61]{Okayama University, Department of Physics, Okayama 700-8530, Japan}
\affiliation[62]{NASA Goddard Space Flight Center, Greenbelt, MD 20771, USA}

\emailAdd{giardiellos@cardiff.ac.uk
}

\abstract{Systematic effects can hinder the sought-after detection of primordial gravitational waves, impacting the reconstruction of the $B$-mode polarization signal which they generate in the cosmic microwave background (CMB). In this work, we study the impact of an imperfect knowledge of the instrument bandpasses on the estimate of the tensor-to-scalar ratio $r$ in the context of the next-generation \textit{LiteBIRD} satellite. We develop a pipeline to integrate over the bandpass transmission in both the time-ordered data (TOD) and the map-making processing steps. We introduce the systematic effect by having a mismatch between the ``real'', high resolution bandpass $\tau$, entering the TOD, and the estimated one $\tau_s$, used in the map-making. We focus on two aspects: the effect of degrading the $\tau_s$ resolution, and the addition of a Gaussian error $\sigma$ to $\tau_s$. To reduce the computational load of the analysis, the two effects are explored separately, for three representative \textit{LiteBIRD} channels (40 GHz, 140 GHz and 402 GHz) and for three bandpass shapes. Computing the amount of bias on $r$, $\Delta r$, caused by these effects on a single channel, we find that a resolution $\lesssim 1.5$ GHz and $\sigma \lesssim 0.0089$ do not exceed the \textit{LiteBIRD} budget allocation per systematic effect, $\Delta r < 6.5 \times 10^{-6}$. We then check that propagating separately the uncertainties due to a resolution of 1 GHz and a measurement error with $\sigma = 0.0089$ in all \textit{LiteBIRD} frequency channels, for the most pessimistic bandpass shape of the three considered, still produces a $\Delta r < 6.5 \times 10^{-6}$. This is done both with the simple deprojection approach and with a blind component separation technique, the Needlet Internal Linear Combination (NILC). Due to the effectiveness of NILC in cleaning the systematic residuals, we have tested that the requirement on $\sigma$ can be relaxed to $\sigma \lesssim 0.05$.}

\begin{document}
\maketitle
\flushbottom

\section{Introduction}
\label{sec:intro}
Measurements of the cosmic microwave background (CMB) have shaped our picture of the standard cosmological model, from the quantum mechanical origin of the Universe to its current energy composition. Several experiments have provided stringent constraints on cosmological parameters~\citep[e.g.,][]{hinshaw2012,Planck2016-l01,Ade:2018,Aiola_2020,Bianchini_2020,P_A_R_Ade_2014} from the observation of the CMB anisotropies in temperature and $E$-mode polarization. The observed $B$-mode polarization is still dominated by measurements of the gravitational lensing term, converting $E$ modes into $B$ modes, and Galactic foreground emission~\citep{planck2016-l04,BICEP2:2015nss}. 
Next generation CMB experiments, such as sets of telescopes part of the ground-based Simons Observatory (SO) \citep{SO} and CMB-S4 \citep{Abazajian:2020dmr}, and the \textit{LiteBIRD} satellite mission \citep{LBspie2024}, have been designed to deliver high-precision measurements of the large-scale $B$ modes which could lead to a high-significance first detection of the primordial signal. This would constrain the amplitude of primordial gravitational waves, parameterized in terms of the tensor-to-scalar ratio $r$. The detection of $r$ is considered the smoking-gun of the inflation paradigm and would revolutionize our understanding of extremely-high-energy physics at play at the beginning of the Universe. So far, we only have upper limits on $r$; the most stringent bound is $r<0.032$ at 95\% CL, based on a re-analysis of BICEP/Keck and \textit{Planck} data~\citep{Tristram:2021tvh}. Improving this upper bound by an order of magnitude would rule out most of the simplest and most compelling early Universe models. 
These next-generation experiments have set ambitious sensitivity goals for $r$: $\sigma(r)\simeq0.002$ from SO, $r<0.001$ at 95\% CL from CMB-S4, and $\sigma(r)\simeq0.001$ from \textit{LiteBIRD}, thus requiring extraordinary control over systematic effects and noise contamination \citep{litebird_ptep, SO}. Residual systematic effects can propagate across the data reduction pipeline and perturb the reconstructed statistics of the CMB by mimicking genuine physical effects of cosmological origin.
If not accounted for, such instrumental systematic effects can be a source of bias in cosmological analyses.

In this work, we consider the effect of bandpass uncertainties, i.e.\ uncertainties in the transmission response of the instrument, on the estimate of $r$. We perform this analysis in the context of the \textit{LiteBIRD} satellite. Specifically, we derive the  bandpass resolution (i.e.\ the resolution used to sample the transmission response) needed to achieve \textit{LiteBIRD}'s scientific goal on $r$.  After that, we obtain requirements for \textit{LiteBIRD} on the bandpass measurement errors, modeled as a Gaussian distributed deviation with respect to the underlying bandpass function. We propagate the effects of the measured bandpass resolution and measurement error to a bias in our estimate of $r$, $\Delta r$. We finally compare our $\Delta r$ with  the \textit{LiteBIRD} budget allocated for each systematic effect of $1 \%$ of the targeted statistical uncertainty ($6.5 \times 10^{-4}$), i.e.\ $\Delta r < 6.5 \times 10^{-6}$~\cite{Mousset_prep, litebird_ptep}.

To reduce the computational cost of the analysis, we divide it in two parts. 
\begin{enumerate}
\itemsep0em
    \item We first consider just three \textit{LiteBIRD} reference channels: the $140$\,GHz channel of the Middle Frequency Telescope (MFT) relevant for the CMB extraction, and the two frequency extremes which are relevant for the astrophysical foreground removal, the $40$\,GHz channel of the Low Frequency Telescope (LFT) and the $402$\,GHz channel of the High Frequency Telescope (HFT). For each of them, we compute the residual maps and corresponding $\Delta r$ induced by assuming different bandpass resolutions and measurement errors $\sigma$. We perform this analysis for three different bandpass shapes: a top-hat with shoulders, and two bandpass profiles, described by Chebyshev filters of order 3 and 5, respectively~\cite{Pozar:882338}. The goal of this first part is to derive the most stringent values of sampling resolution and uncertainty $\sigma$ for the three channels and bandpass shapes considered. 
\item In the second part of the analysis, we fix the Chebyshev profile of order 3 to be the reference bandpass shape (as motivated below). We then propagate separately the systematic effects due to bandpass sampling resolution and measurement uncertainties -- with the values derived in the first step -- across all \textit{LiteBIRD} frequency channels. 
Finally, we estimate the $\Delta r$ induced by the residual systematic perturbation in the final CMB solution.  
\end{enumerate}

We use both a simple deprojection technique (adopted in the first part of the analysis) and a realistic component separation pipeline. This allows us to check that the requirements obtained in the first simplified analysis hold even considering a more realistic framework and the whole instrument.

The paper is structured as follows: Section~\ref{sec:TODformalism} describes our formalism, the bandpass assumptions are introduced in Section~\ref{sec:bandshape}, Sections~\ref{sec:resolution} and \ref{sec:error} present the derivation of requirements for bandpass sampling resolution and measurement uncertainty, while in Section~\ref{sec:drtotal} we assess the robustness of the requirements derived in the previous sections when systematic effects are present in all \textit{LiteBIRD} frequency channels.

\section{TOD and map-making formalism} \label{sec:TODformalism}

To evaluate the effect of different bandpass designs on the $r$ estimate from \textit{LiteBIRD}, we perform a Time Ordered Data (TOD) analysis. We follow the procedure presented in \cite{Giardiello:2021uxq}, which has been ported in the \textit{LiteBIRD} simulation framework \texttt{litebird\_sim}\footnote{\url{https://github.com/litebird/litebird_sim/blob/master/litebird_sim/hwp_sys/hwp_sys.py}, version 0.12.0}~\cite{lbs_prep}.
The TOD for a detector at the time $i$ is computed as:
\begin{equation}\label{eq:dobsnu}
\begin{split} 
\normalsize
d_{\rm{obs}}(t_{i})\,=\, \sum_{X = T, Q, U} \frac{\int d\nu\,K_{\nu}\,\tau(\nu)\,M_{i}^{TX}(\nu)\left(m^X_\mathrm{CMB}+m^X_\mathrm{FG}(\nu\right))}{\int  d\nu K_{\nu} \,\tau (\nu)},
\end{split}
\end{equation}
where $K_{\nu} = \left.\frac{\partial B_{\nu}(T)}{\partial T}\right|_{T_{CMB}}$ is the conversion factor from CMB thermodynamic units to brightness\footnote{$K_{\nu} = \left.\frac{\partial B_{\nu}(T)}{\partial T}\right|_{T_{CMB}} = \left( \frac{2 k_B}{c^2} \right) \nu^2 \frac{x^2 e^x}{(e^x-1)^2}$, where $B_{\nu}(T)$ is Planck's law and $x\equiv h\nu/(k_{B}T_{CMB})$.}, $\tau(\nu)$ is the bandpass, $M_{i}^{TX}(\nu)$ the Mueller matrix of a realistic rotating HWP, encoding also the instrument scanning strategy (for details on its derivation, see \cite{Giardiello:2021uxq}) and $m^X_\mathrm{CMB/FG}$ is the CMB/foreground map for the $X = T, Q, U$ field. These input maps are already smoothed by the Gaussian beam\footnote{We assume an achromatic Gaussian beam, therefore circular and frequency independent, which ensures that the HWP and beam effects are fully decoupled. We refer to Ref.~\cite{Duivenvoorden:2020xzm} for an analysis of the coupling between non-ideal HWP and beams.} of the corresponding channel, see Table 13 of Ref.~\citep{litebird_ptep} for their FWHM.
Similarly to what has been done in \cite{Giardiello:2021uxq}, we simulate the TOD of just two orthogonal detectors\footnote{We select detectors at the boresight or as closest as possible to the boresight for each \textit{LiteBIRD} channel, given the fact that the formalism developed in \cite{Giardiello:2021uxq} assumes orthogonal incidence of light.} for each channel and for one year of observation, in order to reduce the computational load of the analysis. To simulate the Galactic foreground signal, we consider the \texttt{d0s0 PySM}~\cite{Zonca:2021row, Thorne:2016ifb, Pan-ExperimentGalacticScienceGroup:2025vcd} models, i.e.\ Galactic dust and synchrotron with isotropic spectral indexes. For the scope of this analysis, we do not care about the spatial variability of foreground emissions and we restrict ourselves to a simpler model. The bandpass uncertainty can still couple in a non-negligible way with the foreground spatial variability~\cite{Chluba:2017rtj, Vacher:2024adb}, but the study of this effect is deferred to a future work. Anyway, blind component separation methods have been shown to be robust against non-linear effects such as those arising from the coupling of variations of foreground parameters with complex bandpasses, as proved in the case of foreground-only driven SEDs mixing in~\cite{Vacher:2024adb}.

\begin{figure}[tp!]
	\centering
	\includegraphics[width=1\textwidth]{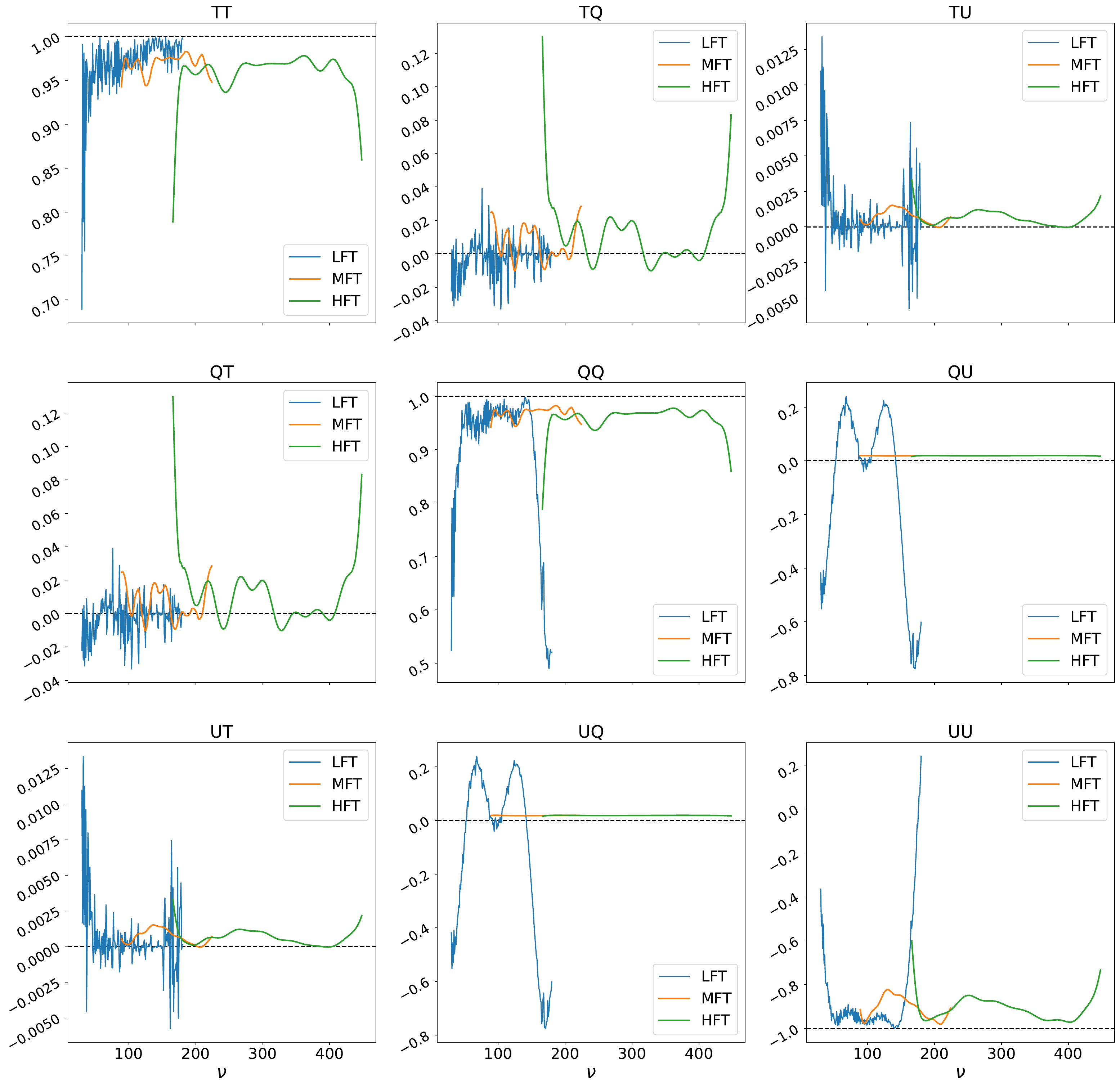}
	\caption{Mueller matrix elements of the LFT, MFT and HFT HWPs as functions of the frequency. These are the components of HWP Mueller matrix in its rest frame. $M_{i}^{TX}(\nu)$ in Eq.~\ref{eq:dobsnu} can be obtained from a combination of these rest-frame components as shown in Ref.~\cite{Giardiello:2021uxq}. The noisy pattern in the LFT HWP elements is caused by its configuration (five-layers sapphire HWP): due to the high refractive index of sapphire, the thickness of the HWP results in fast oscillations of the transmission spectrum. We refer to ~\cite{Giardiello:2021uxq, Monelli:2023wmv} for further details.} \label{fig:Mueller}
\end{figure}

 The denominator in Eq.~\ref{eq:dobsnu} brings the TOD back to CMB temperature units and the assumed bandpass profile is the same as the one at the numerator, assuming perfect dipole calibration. To compute $M^{TX}$ we use the Mueller matrix for realistic \textit{LiteBIRD} HWPs, simulated for the MFT and HFT~\cite{Giardiello:2021uxq} and for the LFT~\cite{Monelli:2023wmv}. The corresponding components of the HWP Mueller matrices in their rest frame are shown in Figure~\ref{fig:Mueller}. We assume a metal-mesh HWP for the Middle and High Frequency Telescopes (MHFT) and a sapphire multilayer HWP for the LFT.

So far, we neglect the contribution of noise in Eq.~\ref{eq:dobsnu} and the TOD are computed as noiseless, to focus only on the systematic effect. The noise contribution (thus the instrument sensitivity) is properly taken into account in the cosmological likelihood as detailed in Section \ref{sec:resolution}. Throughout all the analysis, the noise in the different frequency channels is assumed to be white, i.e.\ non-correlated: $\langle n n^T \rangle = \sigma_n^{2} \mathbb{I}$, where $\sigma_n$ is the uniform noise standard deviation for each channel. The adopted $\sigma_n$ values correspond to those reported in \cite{litebird_ptep}.

To derive a map from the TOD of Eq.~\ref{eq:dobsnu}, we perform the following map-making procedure:

\begin{equation} \label{eq:map-making}
\begin{split}    
&m_{\rm{out}} =\left(\sum_{i} P_{i}^{T} P_{i} \right)^{-1} \left( \sum_{i} P_{i}^{T} d_{\rm{obs}}(t_{i}) \right) \\
&= \left(\sum_{i} P_{i}^{T} P_{i} \right)^{-1} \left( \sum_{i} P_{i}^{T} A_{\mathrm{CMB},i} \, m_\mathrm{CMB}\right) + \left(\sum_{i} P_{i}^{T} P_{i} \right)^{-1}\left(\sum_{i} P_{i}^{T} \int A_{\mathrm{FG},i}(\nu) \, m_\mathrm{FG}(\nu) d\nu \right)\\
\end{split}
\end{equation}
where $i$ represents again the time sample and:
\begin{equation} \label{eq:Acmb}
A_{\mathrm{CMB},i}^X=\,\left(\frac{\int d\nu \,K_{\nu}\,\tau \left(\nu\right)\,M_{i}^{TX}(\nu)}{\int d\nu K_{\nu}\,\tau \left(\nu\right)}\right),
\end{equation}

\begin{equation} \label{eq:Afg}
A_{\mathrm{FG},i}^X=\,\left(\frac{K_{\nu}\,\tau \left(\nu\right)\,M_{i}^{TX}(\nu)}{\int d\nu K_{\nu}\,\tau \left(\nu\right)}\right),
\end{equation}
and $P_{i}$ is the pointing matrix built to recover the CMB component:
\begin{equation} \label{eq:B}
P_{i}^X=\,\left(\frac{\int d\nu \,K_{\nu}\,\tau_{s}\left(\nu\right)\,M_{i}^{TX}(\nu)}{\int d\nu K_{\nu}\,\tau_{s}\left(\nu\right)}\right).
\end{equation}
We have dropped the $X$ for $A$ and $P$ in Eq.~\ref{eq:map-making} and have expressed them in matricial form.

In the map-making we adopt the bandpass profile $\tau_s(\nu)$ (also called ``solver'' bandpass), which represents our estimate of $\tau(\nu)$.\footnote{Bandpasses are measured with a Fourier Transform Spectrometer (FTS)~\citep{bates78}. Typically the FTS is an interferometer with a movable mirror. An example of application to a CMB instrument can be found in Ref.~\cite{Matsuda2019}, where a resolution of 0.5-1 GHz (depending on the exact setting of the instrument) has been achieved with a FTS with dimensions $\sim$1.3 m $\times$ 1 m.} Given a bandpass shape, the TOD is computed using a bandpass $\tau(\nu)$ with high resolution (0.1 GHz),  emulating the real data which is acquired at infinite resolution. Higher resolutions for $\tau(\nu)$ would make the computational cost of Eq.~\ref{eq:dobsnu} too expensive. Instead, the bandpass profile $\tau_s(\nu)$ used in the map-making is computed at a lower resolution by resampling $\tau(\nu)$, with the possibility of introducing Gaussian distributed perturbations, to mimic measurement errors. The results obtained when assuming different sampling resolutions and measurement precisions are presented in Sections \ref{sec:resolution} and \ref{sec:error}. 

The HWP Mueller matrix $M_{i}^{TX}$ assumed in the map-making operation of Eq.~\ref{eq:map-making} is the same used in the TOD, to avoid any mismatch in the estimate of the HWP parameters and to take into account only systematic effects induced by bandpass mismodeling. We refer to Refs.~\cite{Giardiello:2021uxq, Monelli:2023wmv} for discussions on the effects of HWP non-idealities.

All the sky maps are pixelized in the \texttt{HEALPix}~\cite{healpix} format and generated at resolution \texttt{NSIDE=64}, which makes the TOD and map-making computation manageable. Having maps at such a low resolution does not affect the outcome of our analysis, as we aim at assessing the impact of bandpass uncertainties on $r$, which is probed by the very low multipoles ($\ell \lesssim 120$) of the CMB $BB$ power spectrum.

To derive the residual power due to our imperfect estimate of $\tau_s(\nu)$, we build a template map $m_{\rm{templ}}$ using $\tau_s(\nu)$ in both the TOD generation and map-making operation. The residual map $m_{\rm{res}}$ is computed as:
\begin{equation}\label{eq:mres}
\begin{split}    
&m_{\rm{res}} = m_{\rm{out}} - m_{\rm{templ}}  \\
&= \left(\sum_{i} P_{i}^{T} P_{i} \right)^{-1} \left( \sum_{i} P_{i}^{T} A_{\mathrm{CMB},i} \, m_\mathrm{CMB}\right) - m_\mathrm{CMB} \\
&+ \left(\sum_{i} P_{i}^{T} P_{i} \right)^{-1}\left(\sum_{i} P_{i}^{T} \int A_{\mathrm{FG},i}(\nu) \, m_\mathrm{FG}(\nu) d\nu \right) \\
&- \left(\sum_{i} P_{i}^{T} P_{i} \right)^{-1}\left(\sum_{i} P_{i}^{T} \int P_{\mathrm{FG},i}(\nu) \, m_\mathrm{FG}(\nu) d\nu \right), \\
\end{split}
\end{equation}
where $P_{\mathrm{FG}}$ is calculated as in Eq.~\ref{eq:Afg} with $\tau(\nu) = \tau_s(\nu)$.
This residual map is by construction null if $\tau_s(\nu) = \tau(\nu)$ in $m_{\rm{out}}$, i.e.\ if we perfectly recover the true bandpass profile. We are also assuming to perfectly recover the true foreground model (i.e., there is no difference in the foreground input maps used to compute $m_{\rm{out}}$ and $m_{\rm{templ}}$), since we want to just focus on the effect of bandpass uncertainties and not on the combination with biases arising from foreground component separation. 

More details on the formalism can be found in Ref.~\cite{Giardiello:2021uxq}.

\subsection{Bandpass shapes} \label{sec:bandshape}
\begin{figure}[tp!]
	\centering
 {\includegraphics[width=.49\textwidth]{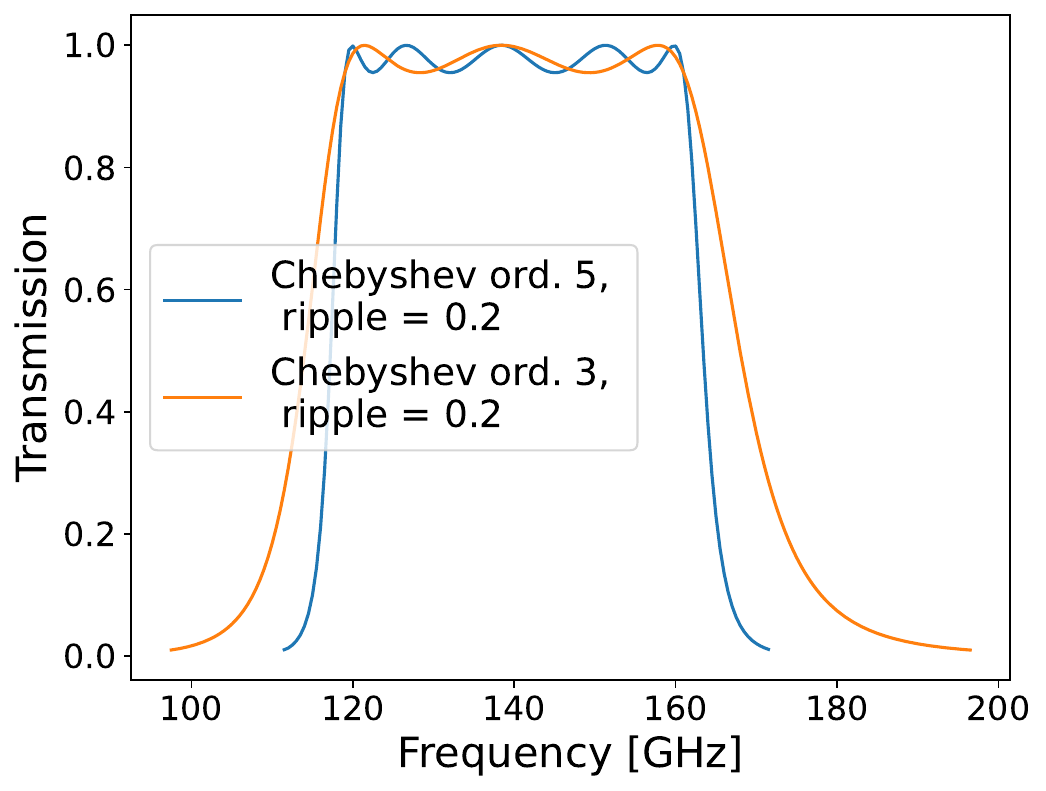}}
 {\includegraphics[width=.49\textwidth]{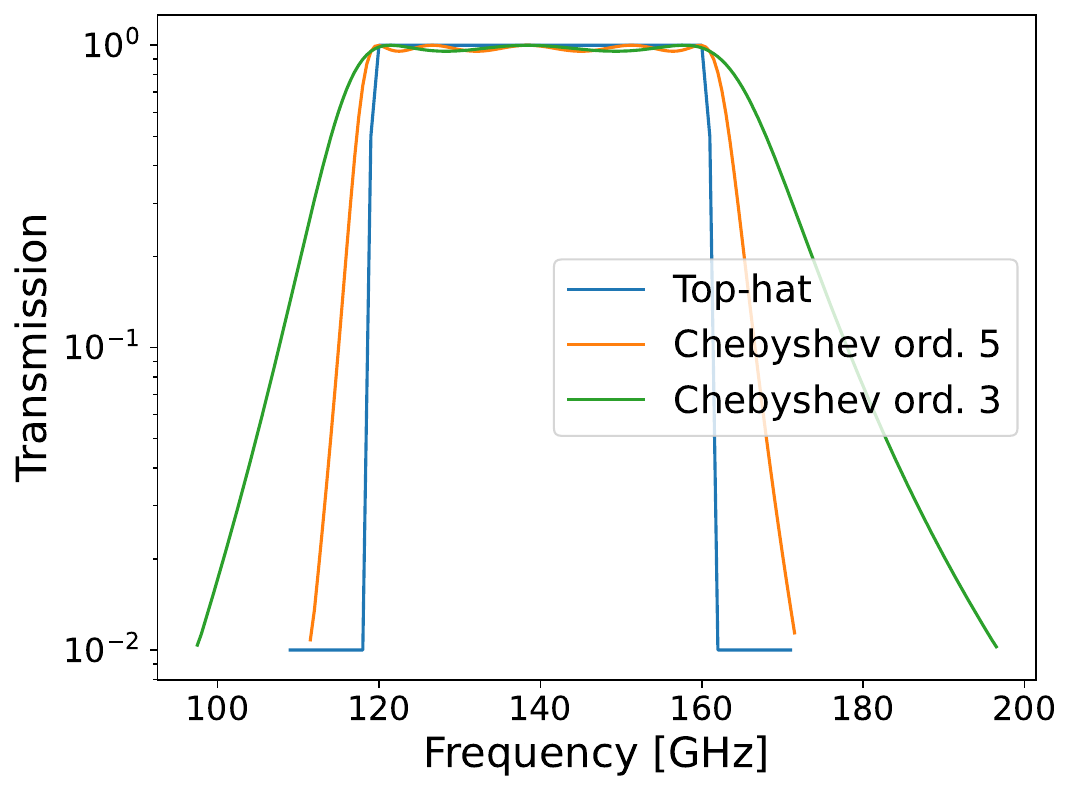}}
	\caption{Left: Bandpass transmissions with a Chebyshev filter of different order parameters for the MFT channel centered at 140 GHz (MFT 140). Right: Bandpass transmissions for the same channel, using different bandpass shapes. The transmission is plotted in log scale to highlight the bandpass tails and the level of the shoulders in the top-hat profile. In both figures the ripple is 0.2 dB.} \label{fig:cheby}
\end{figure}

In the first part of the analysis, we compare systematic residuals in three cases with different bandpass shapes: a simple top-hat with shoulders (despite being not realistic, it still represents an interesting case), a Chebyshev profile with ripple 0.2 dB and order 3 (which is the current design for the on-chip filters of LFT~\cite{LBspie2024}) and the same profile but with order 5 (which is the current design for MHFT). The “ripple” parameter in the Chebyshev profile represents the amount of loss in transmission. The “order” indicates how many oscillations are within the band and it is correlated with the tightness of the bandpass “wings”. A representation of the aforementioned bandpass shapes for the 140 GHz channel can be found in Figure~\ref{fig:cheby}.

We simulate top-hat bandpasses\footnote{See Table 7 of Ref~\cite{litebird_ptep} for their relative bandwidths.} by adding $\pm 10$ GHz to the bandpass edges and assign a level of 0.01 transmission to these out-of-band points. There is a transition point with transmission 0.5 between the in-band and out-of-band points. For Chebyshev bandpasses, we cut them when they reach a level of transmission of 0.01. For an analysis on the effects of out-of-band contamination, we refer to~\cite{Mousset_prep}.

In the second part of the analysis, once we have derived the acceptable sampling resolution and level of measurement error ($\sigma$) for the reference channels (i.e.\ LFT 40, MFT 140 and HFT 402), we fix the bandpass shape. We choose the Chebyshev profile of order 3, since it generally provides the most stringent requirements (see Sections~\ref{sec:resolution} and~\ref{sec:error}), therefore providing an ideal benchmark for assessing the robustness of the derived requirements when bandpass miscalibration is introduced in all frequency channels.\footnote{For a list of the \textit{LiteBIRD} channels, see Table 3 of Ref.\ \cite{litebird_ptep}.}

\section{Requirement on the bandpass sampling resolution} \label{sec:resolution}

In this section, we compare the residuals corresponding to different resolutions of the $\tau_s$ bandpass used in the map-making. We perform this comparison for the three bandpass shapes discussed in Section \ref{sec:bandshape} and, to reduce the computational load, only for three representative \textit{LiteBIRD} channels: the MFT channel centered at 140 GHz (MFT 140), the LFT channel centered at 40 GHz (LFT 40) and the HFT channel centered at 402 GHz (HFT 402). In this way, we can assess the impact of bandpass mismodeling on reconstructed frequency maps which retain great constraining power either in the CMB reconstruction (MFT 140) or in constraining synchrotron (LFT 40) and dust (HFT 402) emission.
We then compute the $\Delta r$ associated with the residuals in each case and derive a requirement on the sampling resolution which allows to meet \textit{LiteBIRD}'s systematic error budget. This is then used in Section \ref{sec:drtotal} for the second part of the analysis, where we verify whether the resolution requirement is still valid even when propagating the mismodeling to all \textit{LiteBIRD} channels. 

\subsection{Derivation of residual power spectra and $\Delta r$ for each channel} \label{subsec:res_cl}
Once a residual map $m^j_{\rm{res}}$ for channel $j$ is derived as detailed in Eq. \ref{eq:mres}, we compute the corresponding $BB$ power spectrum through the \texttt{anafast} routine in the \texttt{healpy} Python package:
\begin{equation}\label{eq:clres}
\begin{split}    
C^{BB,\rm{res},j}_{\ell} = \frac{C^{BB}_{\ell}((m^j_{\rm{res}} \cdot m_{\rm{mask}}) \cdot w_j)}{f_{\rm{sky}} (B^{j}_{\ell} p_{\ell})^2},
\end{split}
\end{equation}
where $m_{\rm{mask}}$ is a customary Galactic mask\footnote{\url{https://irsa.ipac.caltech.edu/data/Planck/release_2/ancillary-data/previews/HFI_Mask_GalPlane-apo0_2048_R2.00/index.html}}, obtained from HFI data processing, with sky fraction $f_{\rm{sky}}=70\%$ (enough for the sky complexity we assume), $B^{j}_{\ell}$ is the beam window function for channel $j$, $p_{\ell}$ is the pixel window function and $w_j$ is the weight for channel $j$, coming from the parametric component separation method \texttt{FGBuster}~\cite{fgbuster, Stompor:2008sf} applied to \textit{LiteBIRD} simulated maps which assume the \texttt{d0s0} foreground model and no systematic effect.\footnote{In our simple approach, using the weights from a parametric component separation method ignoring the systematic effects is the fastest way to weigh the frequency channels. A more correct approach would be to recompute the component separation weights each time for the CMB+foreground maps affected by the systematic effect. In Section~\ref{sec:nilc} we check that our results are robust also to a full component separation method (we considered a blind method). Additionally, in the parametric component separation one could mitigate the systematic effects following the approach of Ref.~\cite{Verges:2020xug}.} The Galactic plane is masked to mimic a realistic data-analysis pipeline of future CMB satellite experiments. Moreover, it allows us to avoid the sky regions where systematic residuals are brighter, being dominated by Galactic foreground emission and color effects as in Eq.~\ref{eq:mres}. Applying the component separation weight $w_j$ allows us to take into account the relative constraining power of channel $j$ in reconstructing the CMB signal. In this way, despite not performing a full component separation procedure, we do not overestimate the residual power coming from extremal channels that are dominated by foreground emissions and would be weighted less in the CMB solution. The $B$-mode angular power spectrum in Eq. \ref{eq:clres}, being computed from $QU$ Stokes parameters maps, may be affected by $E$-to-$B$ leakage in the CMB contribution~\cite{Lewis:2001hp}. However, this potential bias does not affect the outcome of our analysis as the major contribution to $C^{BB,\rm{res},j}_{\ell}$ is coming from the foreground components for all the considered frequency channels, and their $E$-to-$B$ leakage is not as relevant as in the case of CMB dominated maps.

From the residual power spectrum of Eq.~\ref{eq:clres} we can estimate the $\Delta r$ associated with the systematic effect for each frequency channel separately.

To get the bias $\Delta r$ associated to $C^{BB,\rm{res},j}_{\ell}$, we first compute the $BB$ power spectrum for that channel:
\begin{equation}\label{eq:clBB}  
\tilde{C}^{BB,j}_{\ell} = C^{BB, \rm{fid}}_{\ell} + C^{BB, \rm{res}, j}_{\ell} + C^{BB, \rm{noise}}_{\ell},
\end{equation}
where $C^{BB, \rm{fid}}_{\ell}\equiv C_{\ell}^{BB,\rm{lensed}}$ is the fiducial $BB$ power spectrum\footnote{The fiducial power spectrum is computed with \textit{Planck} 2018 fiducial cosmology and no tensor modes.} (lensing only, $r=0$) and $ C_{\ell}^{BB,\rm{noise}}$ is the noise power spectrum of the \textit{LiteBIRD} CMB reconstruction, obtained by combining simulated noise maps in the different frequency channels (see Section~\ref{sec:TODformalism}) with the assumed frequency-dependent component separation weights $w_j$. The residual $C_{\ell}^{BB, \rm{res},j}$ is treated as if it were a spurious cosmological signal leading to a bias in the estimate of $r$.

Such a bias is the best-fit $r$ value obtained assuming the observed power spectrum follows an inverse-Wishart distribution \citep{Hamimeche, Gerbino:2019okg}:

\begin{equation} \label{eq:L}
		\begin{split}
			&-2 \text{ln}\mathcal{\tilde{L}}(r) = -2 \text{ln}\mathcal{L}(\tilde{C}^{BB, j}_{\ell}|C^{BB}_{\ell}(r)+ C_{\ell}^{BB, \rm{noise}}) \\
   & = f_{\rm{sky}} \sum_{\ell} (2 \ell+1) \left[ \frac{\tilde{C}^{BB, j}_{\ell}}{C^{BB}_{\ell}(r)+ C_{\ell}^{BB, \rm{noise}}} - \text{ln} \left(\frac{\tilde{C}^{BB, j}_{\ell}}{C^{BB}_{\ell}(r)+ C_{\ell}^{BB, \rm{noise}}} \right)  \right] \,,
		\end{split}
	\end{equation}
where $\tilde{C}^{BB, j}_\ell$ is the observed power spectrum of Eq.~\ref{eq:clBB}, and $C^{BB}_{\ell}(r) = C_{\ell}^{BB, \rm{lensed}} + C_{\ell}^{BB, \rm{tens}}(r)$ is the theoretical $BB$ power spectrum combining the lensing term and the tensor component for a given value of $r$. The likelihood analysis is restricted to the multipole range\footnote{We computed the spectra up to $3\times\texttt{NSIDE}-1$, with $\texttt{NSIDE}=64$.} $2 \leq \ell \leq 191$ of interest for \textit{LiteBIRD}~\citep{Hazumi:2021yqq}. We renormalize the likelihood at the peak of the distribution and adopt a flat prior on $r$.

The procedure above corresponds to the case where the bandpass uncertainty affects only the reconstruction of a specific channel, while being null in all the others. Although not fully realistic, in this way we can easily estimate the bias on $r$ due to the systematic residual, $C^{BB,\rm{res},j}$, in the considered frequency channel  and then forecast the amount of residuals allowed to meet the budget of $\Delta r < 6.5 \times 10^{-6}$. As discussed above, in this first part of the analysis we deploy this approach only to three reference channels; we then verify in Section~\ref{sec:totaldr_res} that using the same resolution for all channels leads to a similar level of residuals and to $\Delta r < 6.5 \times 10^{-6}$. A more conservative approach would have been to set the threshold per channel to $\Delta r < 6.5 \times 10^{-6}$/$N_{\rm{channels}}$, assuming that each channel contributes equally to the total systematic residual. What we find is that the dominant contribution to the systematic budget comes from the most extreme channels, so this conservative choice would have resulted in too stringent requirements. Our choice is further justified a posteriori by the findings in Section~\ref{sec:drtotal}.

\begin{figure}[tp!]
	\centering
    {\includegraphics[width=.49\textwidth]{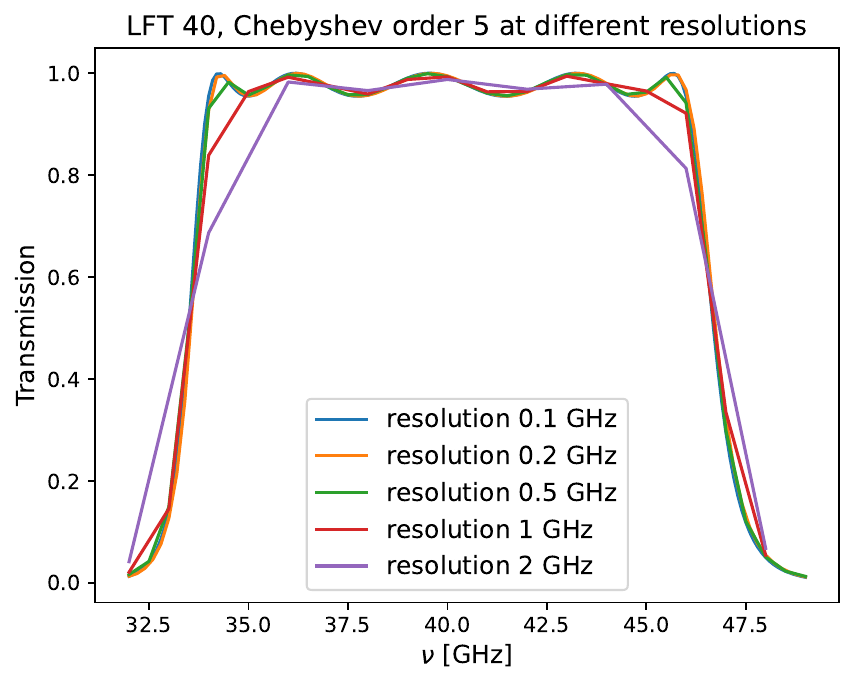}}
	{\includegraphics[width=.49\textwidth]{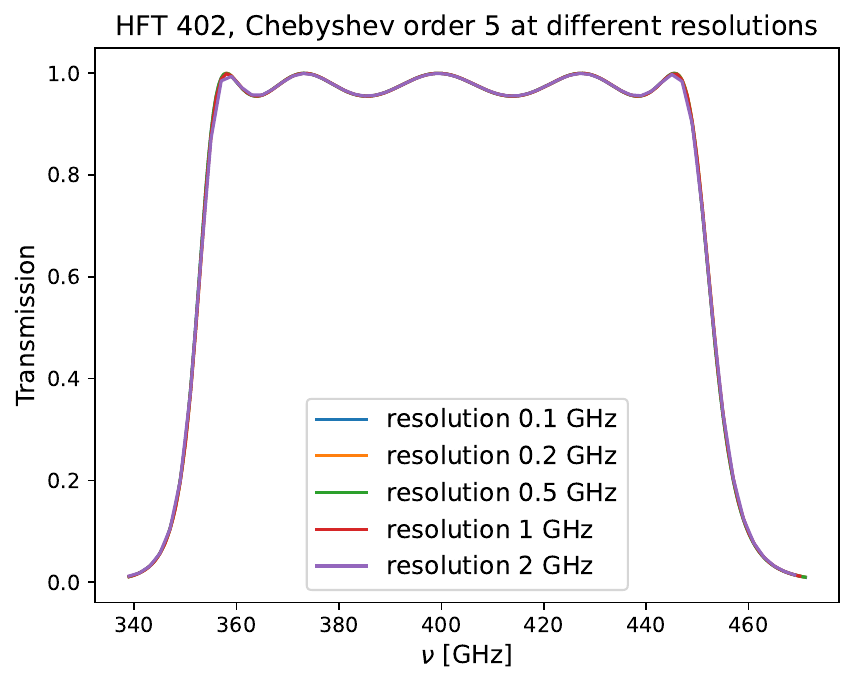}}
	\caption{Chebyshev order 5 bandpass profiles for the LFT 40~GHz channel (left) and the HFT 402~GHz one (right) reconstructed with different resolutions. The coarser the resolution, the worse is the reconstruction of the bandpass shape (which becomes more stable with finer resolutions of the order of a few tenths of GHz), especially for narrower bandwidths and more complicated shapes such as the Chebyshev ones. The effect is clearly worse in the left panel compared to the right one, since the bandwidth of channel LFT 40 is the narrowest.} \label{fig:lft40_band}
\end{figure}

\begin{figure}[tp!]
	\centering
    {\includegraphics[width=.49\textwidth]{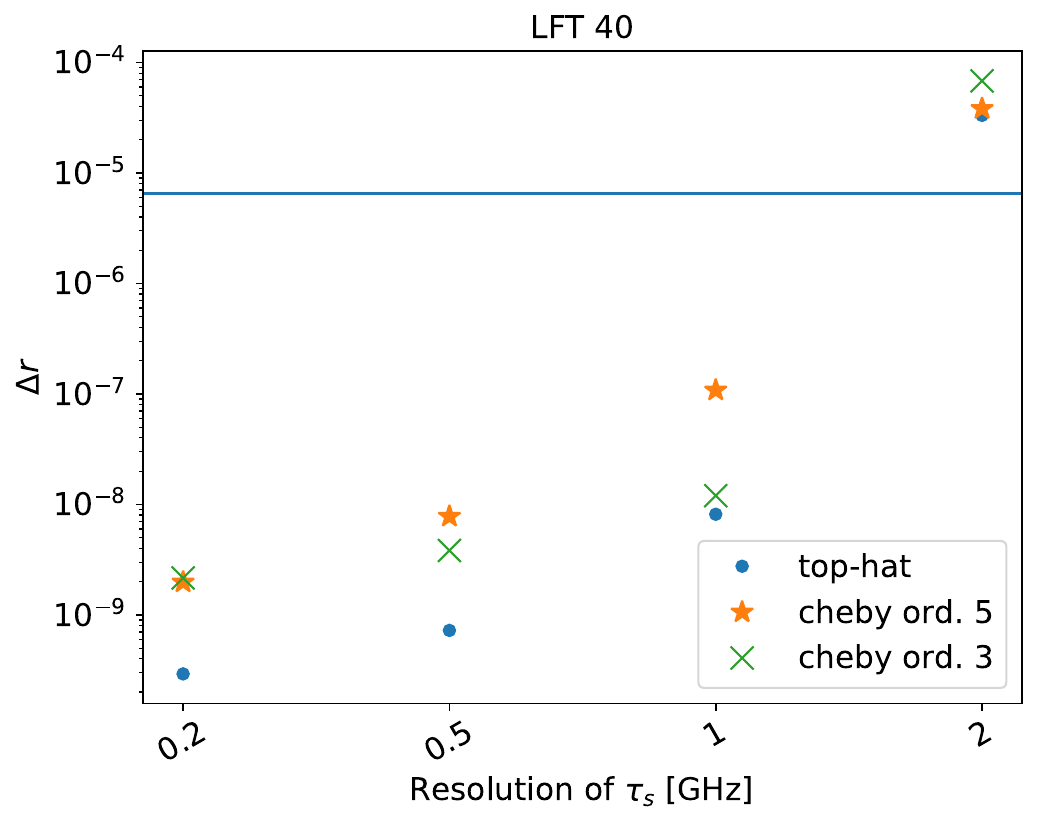}} 
    {\includegraphics[width=.49\textwidth]{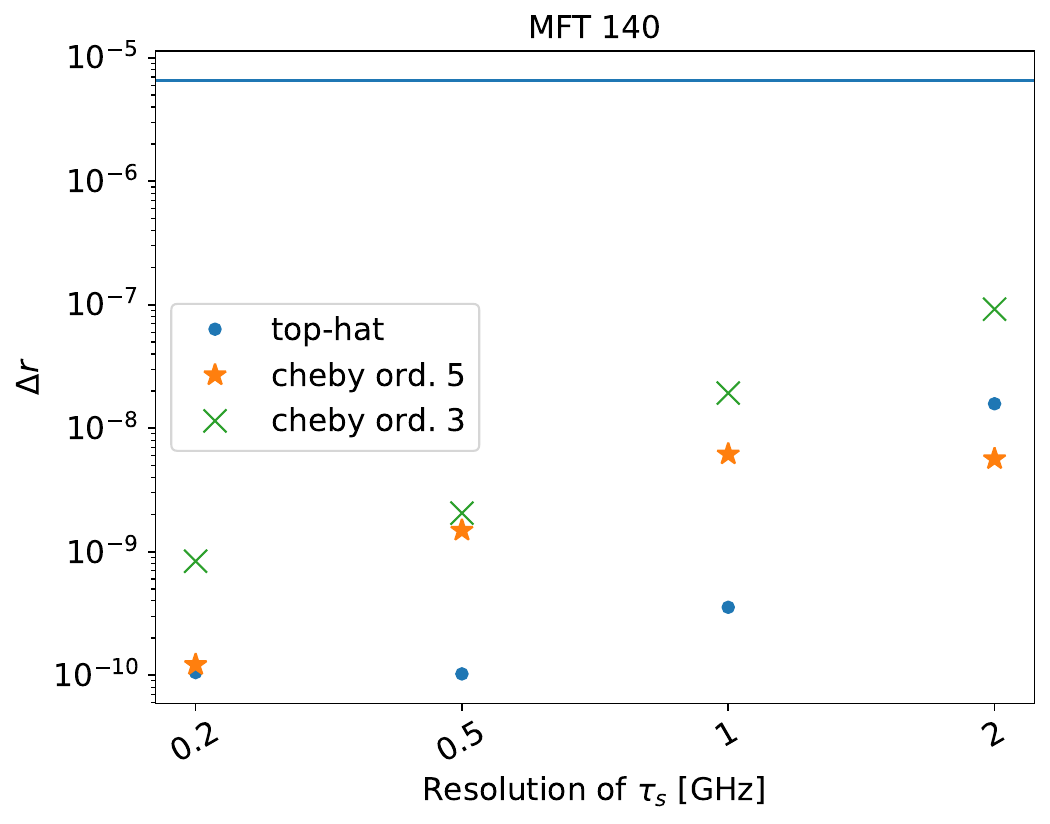}}
    {\includegraphics[width=.49\textwidth]{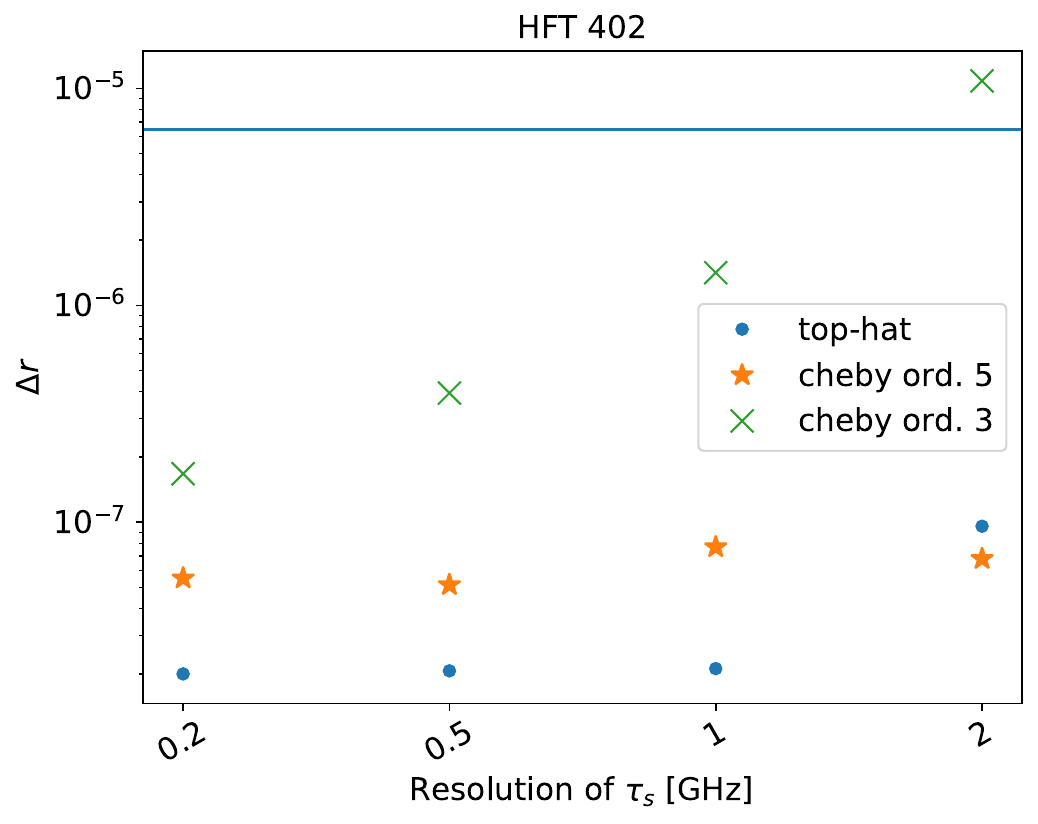}}
	\caption{$\Delta r$ as function of different values of $\tau_s$ resolution for the three reference channels considered. We use blue dots for top-hat bandpass profiles, orange stars for Chebyshev order 5  and green crosses for Chebyshev order 3. We conclude that, as a single requirement for all channels, a resolution $\lesssim 1.5$ GHz is needed not to exceed the threshold $\Delta r = 6.5 \times 10^{-6}$ (marked by the blue horizontal line). } \label{fig:dr_resolution}
\end{figure}

We derive $\Delta r$ for $40,\ 140$ and $402$ GHz channels, three bandpass shapes and different values of bandpass resolution (see Figure~\ref{fig:dr_resolution} and Table~\ref{tab:dr_res}). We consider sampling resolutions of 0.2 GHz, 0.5 GHz, 1 GHz and 2 GHz. As expected, we observe that the coarser the resolution, the worse is the reconstruction of the bandpass shape, especially for the Chebyshev shapes and for bandpasses with narrower bandwidth (see Figure \ref{fig:lft40_band}).

\begin{itemize}
    \item {\bf LFT 40:} For the LFT 40~GHz channel, the systematic residuals exceed the limit for resolutions $\gtrsim 1.5$ GHz for all the three bandpass shapes. The amplitude of residuals is large because lowering the sampling resolution for channels with small bandwidth results in a poor bandpass reconstruction; see Figure~\ref{fig:lft40_band}. Moreover, the amplitude of the perturbation is between 1 and 2 orders of magnitude larger than in the case of an ideal HWP, since the simulated LFT HWP Mueller matrix is quite far from ideality for the lowest LFT channel (see Figure \ref{fig:Mueller}).
    
    \item {\bf MFT 140:} For the MFT 140~GHz channel, all the assumed resolutions seem appropriate, for all bandpass shapes (see Figure \ref{fig:dr_resolution}). The residuals due to degrading resolutions of $\tau_s$ do not exceed the limit $\Delta r < 6.5 \times 10^{-6}$. This is due to the fact that systematic residuals, being dominated by foreground distortions (see Eq. \ref{eq:mres}), are relatively less dominant in CMB channels. We repeated the same analysis in the case of an ideal HWP, and the results are very similar to the non-ideal HWP case. In fact, the Mueller matrix elements assumed for MFT are not as far from ideality as, e.g.\  the LFT ones.

   \item {\bf HFT 402:} For the HFT 402~GHz channel, the residuals for top-hat and Chebyshev order 5 shapes allow to meet the $\Delta r$ requirement for all the selected resolutions. In the case of Chebyshev order 3 profile, we obtain, instead, $\Delta r > 6.5 \times 10^{-6}$ for resolution > 1.5 GHz. As it is found for the low-frequency case, for this frequency channel the systematic effect is larger than in the case of an ideal HWP, due to the deviations of the HFT HWP from ideality, see Figure \ref{fig:Mueller}.

\end{itemize}

By combining the results obtained in the three different channels, we can conclude that a sampling resolution $\lesssim 1.5$ GHz is enough to meet the requirement on $\Delta r$ assigned to this systematic effect. 

	\begin{table*}
		\begin{center} 
			\caption{Residual $\Delta r$ for the three test channels, the selected bandpass shapes and the different resolutions. The values in red are higher than the threshold $\Delta r < 6.5 \times 10^{-6}$.
			}  \label{tab:dr_res}
			\begin{tabular}{ P{3 cm}|| P{2 cm}|P{2 cm}|P{2 cm}|P{2 cm} }
				\hline
				\hline
				Resolution & 0.2 GHz & 0.5 GHz &  1 GHz & 2 GHz \\
\hline
\hline
			{\makecell{$\Delta r $ LFT 40, \\  top-hat}} & $2.9 \times 10^{-10}$ & $7.2 \times 10^{-10}$ & $8.1 \times 10^{-09}$ & ${ \bf \color{red} 3.3 \times 10^{-05}}$ \\
				\hline
			{\makecell{$\Delta r $ LFT 40, \\  Cheby. order 5}}  &$2.0 \times 10^{-09}$ & $7.8 \times 10^{-09}$ & $1.1 \times 10^{-07}$ & ${ \bf \color{red} 3.8 \times 10^{-05}}$ \\
				\hline
			{\makecell{$\Delta r $ LFT 40, \\  Cheby. order 3}} &$2.2 \times 10^{-09}$ & $3.8 \times 10^{-09}$ & $1.2 \times 10^{-08}$ & ${ \bf \color{red} \color{red} 6.8 \times 10^{-05}}$ \\
\hline
\hline
            {\makecell{$\Delta r $ MFT 140, \\  top-hat}} & $1.1 \times 10^{-10}$ & $1.0 \times 10^{-10}$ & $3.6 \times 10^{-10}$ & $1.6 \times 10^{-08}$ \\
				\hline
			{\makecell{$\Delta r $ MFT 140, \\  Cheby. order 5}}  & $1.2 \times 10^{-10}$ & $1.5 \times 10^{-09}$ & $6.2 \times 10^{-09}$ & $5.6 \times 10^{-09}$ \\
				\hline
			{\makecell{$\Delta r $ MFT 140, \\  Cheby. order 3}} & $8.4 \times 10^{-10}$ & $2.1 \times 10^{-09}$ & $1.9 \times 10^{-08}$ & $9.2 \times 10^{-08}$ \\
\hline
\hline
			{\makecell{$\Delta r $ HFT 402, \\  top-hat}} &$2.0 \times 10^{-08}$ & $2.1 \times 10^{-08}$ & $2.1 \times 10^{-08}$ & $9.6 \times 10^{-08}$\\
				\hline
			{\makecell{$\Delta r $ HFT 402, \\  Cheby. order 5}}  & $5.5 \times 10^{-08}$ & $5.2 \times 10^{-08}$ & $7.7 \times 10^{-08}$ & $6.8 \times 10^{-08}$\\
				\hline
			{\makecell{$\Delta r $ HFT 402, \\  Cheby. order 3}} & $1.7 \times 10^{-07}$ & $3.9 \times 10^{-07}$ & $1.4 \times 10^{-06}$ & ${ \bf \color{red}\color{red} 1.1 \times 10^{-05}}$ \\

    \hline
    \hline

				\hline
			\end{tabular}
		\end{center}
	\end{table*}

\section{Requirement on the bandpass measurement error} \label{sec:error}
In this section, we compare the residuals corresponding to different measurement errors in the bandpass $\tau_s$ assumed in the map-making. The analysis pipeline is the same detailed in Section \ref{sec:resolution}. We perturb $\tau_s$ with a Gaussian distributed error with values of $\sigma$ spanning per mille to tens of percent uncertainty. To apply the measurement error, we follow this procedure: we start from the high-resolution $\tau$ profile, we add a Gaussian error with $\sigma_{\rm{input}} =$ [0.005, 0.01, 0.03, 0.05, 0.1], and we then bin the perturbed bandpass to a lower resolution. $\tau_s$ will be this lower resolution and perturbed bandpass. Due to this binning procedure, the effective $\sigma$ is rescaled by a factor $\sqrt{0.1}/\sqrt{(\tau_s \, \text{resolution})}$, where 0.1 GHz is the resolution of $\tau$.

In this part of the analysis, we want to assess the impact of the measurement error and disentangling it from that of the sampling resolution, since we aim at deriving an independent requirement for this effect. To this scope, we set $\tau_s$ to an optimal resolution, such that the two effects can be considered as almost independent. We find this resolution to be 0.5 GHz. In such a case, keeping fixed $\sigma$ and the error realization, we indeed find a balancing of the effects of bandpass shape distortions due to the coarser resolution (see Figure~\ref{fig:lft40_band}) and the smoothing of the error oscillations when degrading the resolution. 
Furthermore, for $\tau_s=0.5$ GHz the residuals are more stable to changes in bandpass error realizations compared to a sampling resolution of 1 GHz.\footnote{This is probably due to the fact that, having less data points at 1 GHz with respect to 0.5 GHz, we have less oscillations and a stronger dependence on the specific bandpass error realization.} We have verified that, at the chosen resolution, the effect of the measurement error is almost decoupled from the effect of undersampling: we have repeated the analysis presented in this section using 0.5 GHz as original resolution for $\tau$, such that the only present systematic effect is due to the measurement errors. The results are compatible with those shown in the following, thus justifying the choice of $\tau_s=0.5$ GHz and of treating the resolution and measurement error effects as if they were independent. We assume no correlation between the bandpass errors of different channels.

\begin{figure}[tp!]
	\centering
    {\includegraphics[width=.8\textwidth]{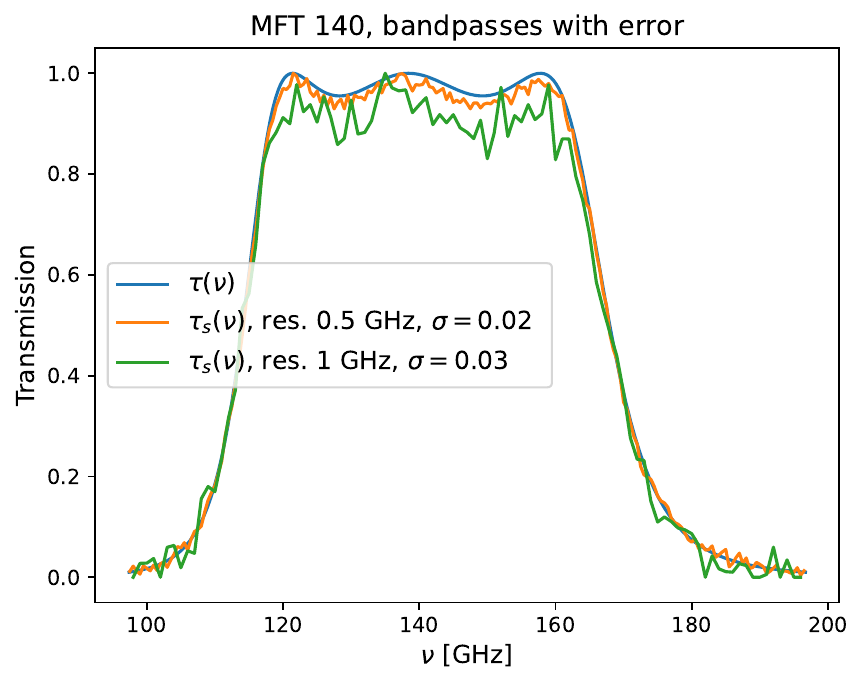}} 
	\caption{Example of bandpasses $\tau_s$ when perturbed by a Gaussian distributed error, with respect to the unperturbed and high resolution $\tau$ (in blue). The bandpasses are renormalized, to have max($\tau_s$) = 1. We show one case with resolution 0.5 GHz and $\sigma = 0.02$ (orange), and one with resolution 1 GHz and $\sigma = 0.03$ (green). They have two different bandpass error realizations. } \label{fig:band_sigma}
\end{figure}

\begin{figure}[tp!]
	\centering
  {\includegraphics[width=.49\textwidth]{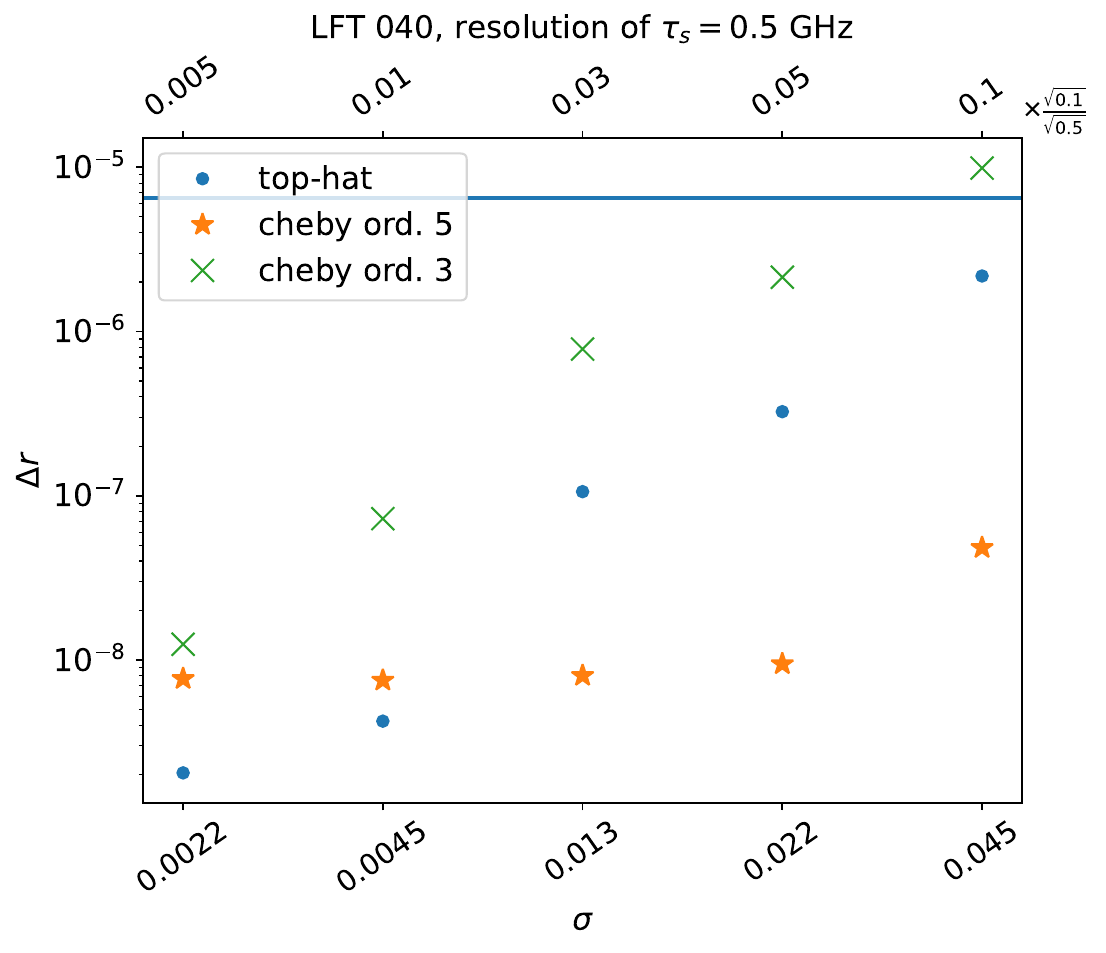}} 
{\includegraphics[width=.49\textwidth]{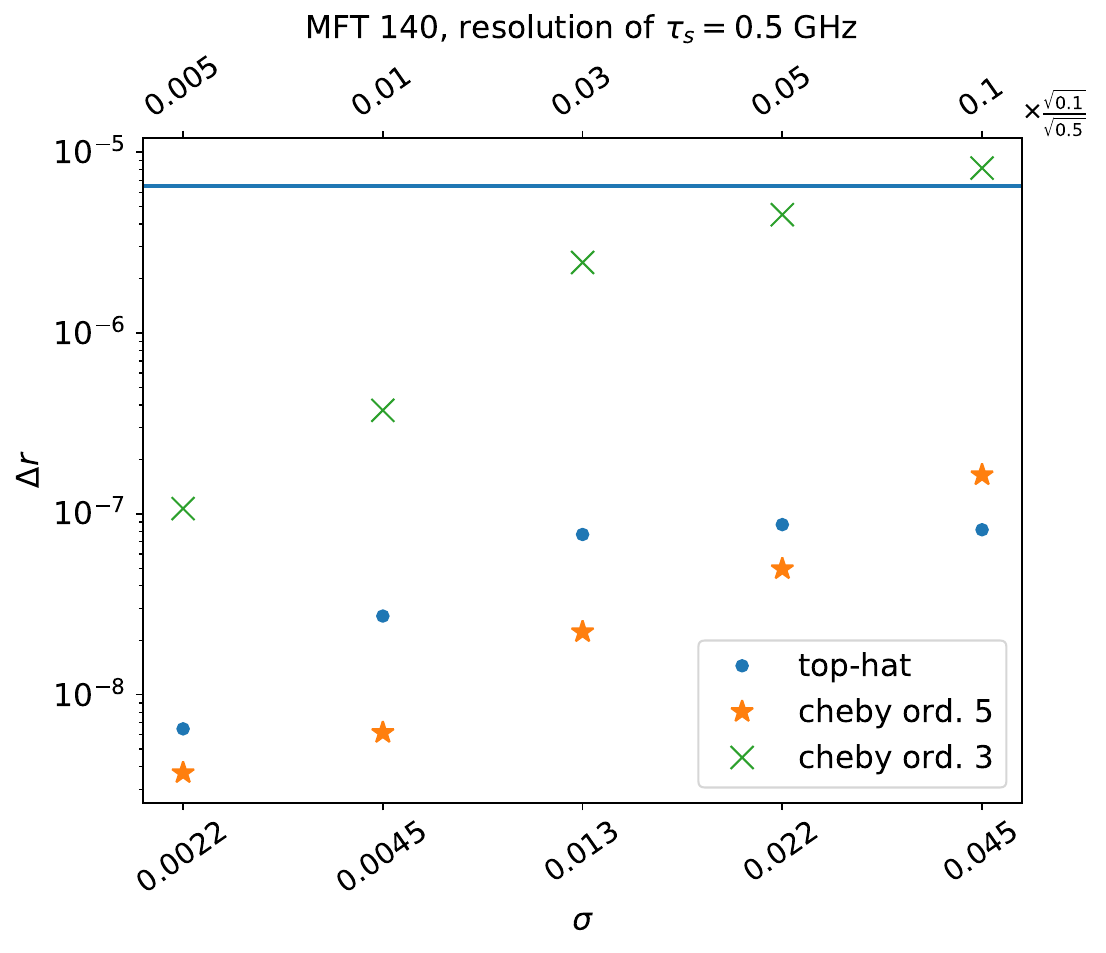}}
{\includegraphics[width=.49\textwidth]{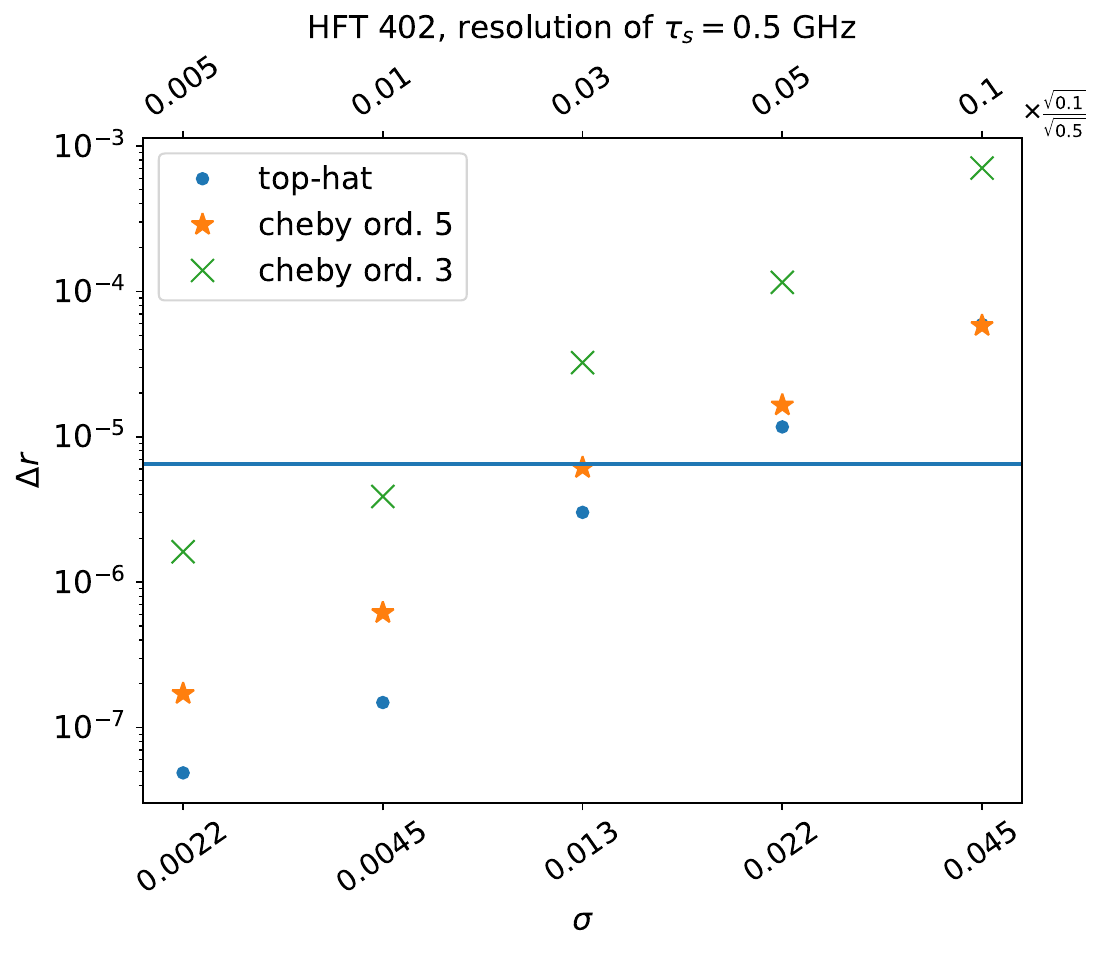}}
	\caption{$\Delta r$ for different values of Gaussian error, $\sigma$, present in $\tau_s$ (which has a resolution of 0.5 GHz), and for the three reference channels. The x-axis at the top shows the $\sigma_{\rm{input}}$ values of the Gaussian error applied to the bandpass with 0.1 GHz resolution, then binned down to 0.5 GHz (see the text for more details). The $\sqrt{0.1}/\sqrt{0.5}$ factor takes into account the rescaling of the error in $\tau_s$ due to the binning; the effective $\sigma$ values are shown in the bottom x-axis. We use blue dots for top-hat profiles, orange stars for Chebyshev order 5 and green crosses for Chebyshev order 3. We conclude that, to satisfy the requirement across all frequencies, a $\sigma \lesssim 0.02 \times \sqrt{0.1}/\sqrt{0.5} = 0.0089$ is needed not to exceed the threshold $\Delta r = 6.5 \times 10^{-6}$ (marked by the blue horizontal line). This plot refers to one specific bandpass error realization, but we have verified that the results are stable to the change of realizations.} \label{fig:dr_sigma}
\end{figure}

The results shown in Figure~\ref{fig:dr_sigma} and Table~\ref{tab:dr_sigma} are just for one bandpass error realization, but we have repeated this test with few other realizations.\footnote{We consider only few realizations for the computational cost of the analysis. When doing NILC component separation in Section~\ref{sec:nilc} we use 50 different CMB + bandpass noise + map noise realizations, so we can test the robustness of the requirements to the different bandpass error realizations.} After having applied the Gaussian distributed errors on the different bandpass profiles, we renormalize them to have max($\tau_s$) = 1 (see Figure~\ref{fig:band_sigma}). We have verified that such a renormalization procedure does not affect our results, since the bandpasses appear always in a ratio (see Eqs. \ref{eq:Acmb}-\ref{eq:B}).

As done in Section~\ref{subsec:res_cl}, we derive $\Delta r$ for the three reference channels and the three bandpass shapes, and explore here the different values of $\sigma$.

\begin{itemize}
    \item {\bf LFT 40} For the LFT 40~GHZ channel, all the residuals for $\sigma < 0.1 \times \sqrt{0.1}/\sqrt{0.5} = 0.045$ are acceptable. Repeating this test with another bandpass error realization, we get higher residuals and an acceptable $\sigma < 0.05  \times \sqrt{0.1}/\sqrt{0.5} = 0.022$. This channel is however not the main driver of a joint requirement on $\sigma$, as discussed below and shown in Figure~\ref{fig:dr_sigma}.
    
    \item {\bf MFT 140} For the MFT 140~GHz channel, the residuals are acceptable for $\sigma < 0.045$, especially for the top-hat and Chebyshev order 5 profiles (see Figure~\ref{fig:dr_sigma}). 

   \item {\bf HFT 402} The HFT 402~GHz channel drives the most stringent constraints on $\sigma$: tests done with different error realizations show that $\sigma \lesssim 0.02  \times \sqrt{0.1}/\sqrt{0.5} = 0.0089$ is needed for this channel.\footnote{This requirement comes from selecting the middle point between $0.01  \times \sqrt{0.1}/\sqrt{0.5}$ and $0.03  \times \sqrt{0.1}/\sqrt{0.5}$ and not from a precise interpolation of the curves in Figure~\ref{fig:dr_sigma}. This approximate approach is used to avoid relying on the interpolation of the results of one single realization.}
\end{itemize}
The requirement on $\sigma$ which is valid for all cases is $\sigma \lesssim 0.02  \times \sqrt{0.1}/\sqrt{0.5} = 0.0089$ (see Figure~\ref{fig:dr_sigma} and Table~\ref{tab:dr_sigma}). This will be used in Section \ref{sec:totaldr_sigma}.
Repeating these tests with a lower resolution (1 GHz) shows higher residuals than what we have seen so far, but in general the constraint of $\sigma \lesssim 0.0089$ appears to still be valid also in this case. 

This requirement is more stringent than the level of measurement statistical error of the POLARBEAR bandpasses \cite{Matsuda2019}. We will verify in Section~\ref{sec:nilc} that this $\sigma$ requirement can be relaxed up to $\sigma \lesssim 0.05$ when using NILC.

	\begin{table*}
		\begin{center} 
			\caption{Residual $\Delta r$ for the three test channels, the selected bandpass shapes and the different values of $\sigma$ for the Gaussian error. The values in red are higher than the threshold $\Delta r < 6.5 \times 10^{-6}$.
			}  \label{tab:dr_sigma}
			\begin{tabular}{ P{2.5 cm}|| P{2 cm}|P{2 cm}|P{2.1 cm}|P{2.1 cm}|P{2.1 cm} }
				\hline
				\hline
				$\sigma$ & 0.0022 & 0.0045 &  0.013 & 0.022 & 0.045 \\
\hline
\hline
              {\makecell{$\Delta r $ LFT 40, \\  top-hat}} & $2.1 \times 10^{-09}$ & $4.2 \times 10^{-09}$ & $1.1 \times 10^{-07}$ & $3.3 \times 10^{-07}$ & $2.2 \times 10^{-06}$\\
				\hline
			{\makecell{$\Delta r $ LFT 40, \\  Cheby. order 5}}  & $7.7 \times 10^{-09}$ & $7.5 \times 10^{-09}$ & $8.0 \times 10^{-09}$ & $9.5 \times 10^{-09}$ & $4.8 \times 10^{-08}$\\
				\hline
			{\makecell{$\Delta r $ LFT 40, \\  Cheby. order 3}} & $1.3 \times 10^{-08}$ & $7.3 \times 10^{-08}$ & $7.8 \times 10^{-07}$ & $2.1 \times 10^{-06}$ & ${ \bf \color{red}\color{red} 9.9 \times 10^{-06}}$\\
\hline
\hline

			{\makecell{$\Delta r $ MFT 140, \\  top-hat}} & $6.5 \times 10^{-09}$ & $2.7 \times 10^{-08}$ & $7.7 \times 10^{-08}$ & $8.7 \times 10^{-08}$ & $8.2 \times 10^{-08}$\\
				\hline
			{\makecell{$\Delta r $ MFT 140, \\  Cheby. order 5}}  &$3.7 \times 10^{-09}$ & $6.2 \times 10^{-09}$ & $2.2 \times 10^{-08}$ & $5.0 \times 10^{-08}$ & $1.6 \times 10^{-07}$
\\
				\hline
			{\makecell{$\Delta r $ MFT 140, \\  Cheby. order 3}} & $1.1 \times 10^{-07}$ & $3.7 \times 10^{-07}$ & $2.5 \times 10^{-06}$ & $4.5 \times 10^{-06}$ & ${ \bf \color{red}\color{red} 8.2 \times 10^{-06}}$\\
\hline
\hline
		{\makecell{$\Delta r $ HFT 402, \\  top-hat}} & $4.9 \times 10^{-08}$ & $1.5 \times 10^{-07}$ & $3.0 \times 10^{-06}$ & ${ \bf \color{red}\color{red} 1.2 \times 10^{-05}}$ & ${ \bf \color{red}\color{red} 5.9 \times 10^{-05}}$\\
				\hline
			{\makecell{$\Delta r $ HFT 402, \\  Cheby. order 5}}  & $1.7 \times 10^{-07}$ & $6.2 \times 10^{-07}$ & $6.1 \times 10^{-06}$ & ${ \bf \color{red}\color{red} 1.7 \times 10^{-05}}$ & ${ \bf \color{red}\color{red} 5.8 \times 10^{-05}}$\\
				\hline
			{\makecell{$\Delta r $ HFT 402, \\  Cheby. order 3}} & $1.6 \times 10^{-06}$ & $3.9 \times 10^{-06}$ & ${ \bf \color{red}\color{red} 3.2 \times 10^{-05}}$ & ${ \bf \color{red}\color{red} 1.2 \times 10^{-04}}$ & ${ \bf \color{red}\color{red} 7.1 \times 10^{-04}}$\\
    \hline
				\hline
			\end{tabular}
		\end{center}
	\end{table*}

\section{Testing the requirements on the full \textit{LiteBIRD} frequency configuration} \label{sec:drtotal}
After deriving the acceptable levels for bandpass shape, resolution and measurement precision from the previous analyses, we proceed to compute the total $\Delta r$ when bandpass mismatches which match the derived requirements are injected in all frequency channels. As before, the residual maps for each channel are combined with $w_j$, the weights for CMB reconstruction obtained from the \texttt{FGBuster}  method and referring to the foreground model \texttt{d0s0}, to mimic a component separation procedure. 

To derive the bias $\Delta r$ due to the residuals from all the frequency channels, we follow the same procedure detailed in Section~\ref{subsec:res_cl}, except for the residual power spectrum which is computed as:
\begin{equation}\label{eq:clres_tot}   
C^{BB,\rm{res}}_{\ell} = C^{BB} (\sum_j (m^j_{\rm{res}} \cdot m_{\rm{mask}}) \cdot w_j )/(f_{\rm{sky}} (B^{\rm{max}}_{\ell} p_{\ell})^2).
\end{equation}
where $B^{\rm{max}}_{\ell}$ is the beam of the channel with the largest FWHM (corresponding to LFT 040, FWHM $\sim 80'$), and all residual maps have been smoothed to this resolution. This is the power spectrum of the co-added residual maps from all the frequency channels, which are appropriately weighted by the corresponding foreground cleaning weights $w_j$.

\subsection{Testing the resolution requirement}    \label{sec:totaldr_res}
In Section \ref{sec:resolution} we find that the sampling resolution of the map-making bandpass $\tau_s$ needs to be $\lesssim$ 1.5 GHz. In this section we consider a resolution of 1 GHz, this being the most frequently assumed sampling resolution, and a bandpass shape corresponding to the Chebyshev order 3 case, for which we get the most stringent requirements. 

\begin{figure}[tp!]
{\includegraphics[width=0.9\textwidth]{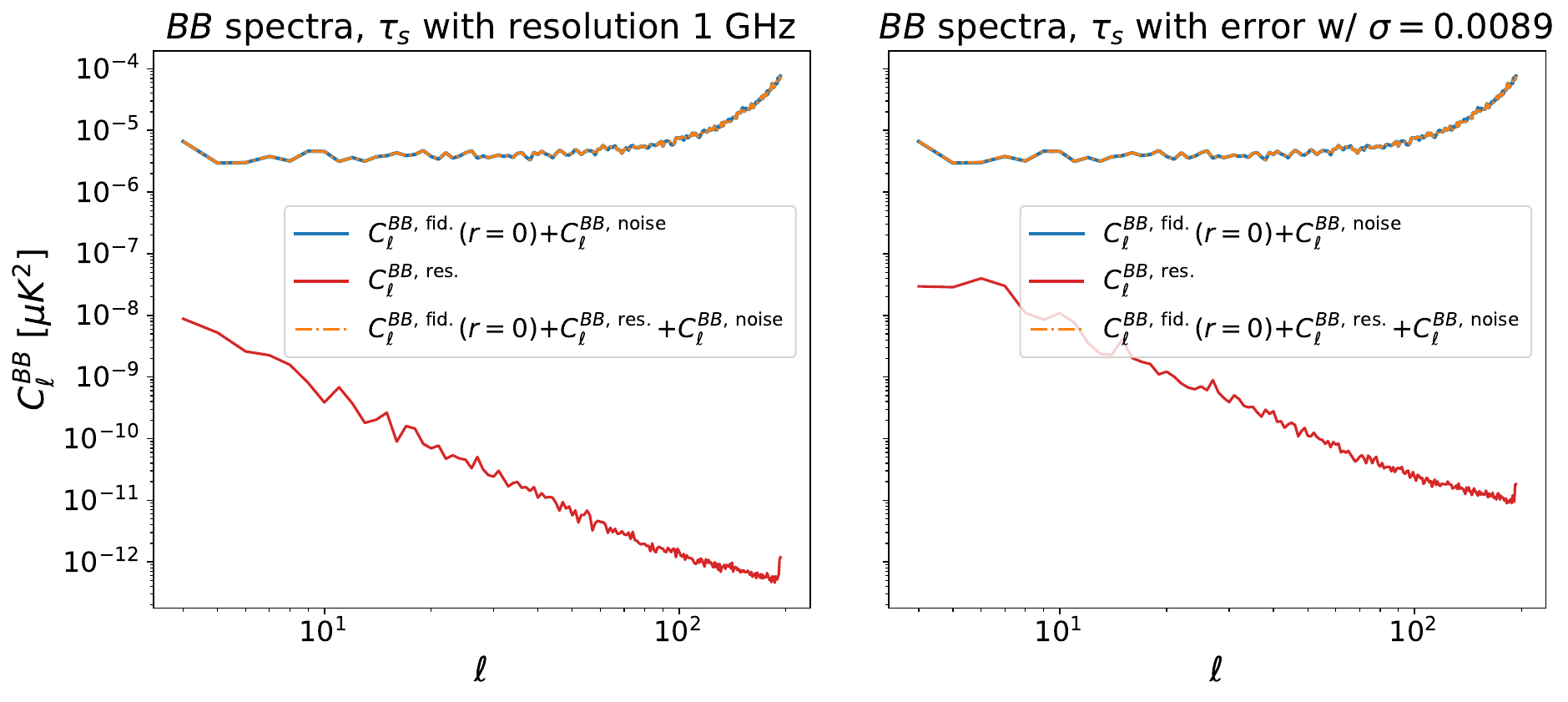}}
\caption{Total residual power spectra when using a resolution of 1 GHz (left) or a resolution of 0.5 GHz and a Gaussian error with $\sigma = 0.0089$ (right) for $\tau_s$ across all \textit{LiteBIRD}'s frequency channels. The blue curve is the fiducial + noise power spectra, the red one is the systematic residual power spectrum and the orange dash-dotted line is the sum of the two.} \label{fig:residuals}
\end{figure}

Applying this setting to all the \textit{LiteBIRD} frequency channels, computing the total residual power spectrum as in Eq.~\ref{eq:clres_tot} (see Figure~\ref{fig:residuals}) and computing the likelihood in Eq.~\ref{eq:L}, we derive a bias $\Delta r = 6.5 \times 10^{-7}$. This is well within the limit assigned to each systematic effect. We can then conclude that a sampling resolution of 1 GHz is acceptable for the instrument across the whole frequency range.

{
\noindent{}
\noindent{}
}

\subsection{Testing the $\sigma$ requirement}    \label{sec:totaldr_sigma}
The requirement found in Section \ref{sec:error} for the standard deviation $\sigma$ of the bandpass Gaussian measurement error is $\sigma \lesssim 0.0089$. As done in the previous sub-section, we now check that the requirement from the first restricted analysis is still satisfied when the perturbation is injected to all frequency channels. We consider only one bandpass shape, the Chebyshev order 3 profile, for which we have the most stringent requirements. The resolution of $\tau_s$ is 0.5 GHz (as explained in Section \ref{sec:error}) and the Gaussian error is applied with $\sigma = 0.0089$. We apply this setting to all \textit{LiteBIRD} channels, using a different bandpass error realization for each channel in order to account for different uncorrelated statistical errors affecting the different bandpass measurements.

Figure~\ref{fig:residuals} shows the residual power spectra for both the case with resolution 1 GHz (described in the previous subsection) and the case with $\sigma \lesssim 0.0089$. The residual in the latter case is higher, which is also reflected in the bias on $r$.

The bias on $r$ derived in this case is $\Delta r = 4.7 \times 10^{-6}$, which is larger than the case with resolution 1 GHz and no error, but still within the limit of $\Delta r = 6.5 \times 10^{-6}$. 
We then conclude that the requirement of $\sigma \lesssim 0.0089$ (using a resolution of 0.5 GHz) is appropriate for the whole \textit{LiteBIRD} instrument. 

\subsection{Testing the requirements with blind component separation} \label{sec:nilc}
As a final validation of our results, we apply an alternative component separation pipeline on the perturbed simulations. We use the Needlet Internal Linear Combination (NILC)~\cite{nilc-basak, Carones:2022dnm}. The NILC technique performs a reconstruction of the CMB signal from the input multifrequency dataset with minimal assumptions. Specifically, it combines the maps with frequency- and pixel-dependent weights which allow to recover the blackbody signal with minimum overall variance. To effectively tackle the different contaminants, this combination is implemented separately at different angular scales by leveraging the framework of needlet filtering~\cite{Baldi:2006dk}. Minimum variance techniques, as NILC, need to be applied on scalar fields, therefore the simulated $Q$ and $U$ maps are converted into $B$ modes through a full-sky harmonic transformation. 

To mimic a realistic data analysis scenario, the component separation step is performed on $m_{\rm{out/templ}} + m_{\rm{noise}}$, i.e.\ we add a white noise map for each channel on top of $m_{\rm{out/templ}}$. We generate $m_{\rm{out}}$ and $m_{\rm{templ}}$ with 50 different CMB and noise realizations to improve the significance of the results. To test the robustness of both requirements derived in Sections \ref{sec:resolution} and \ref{sec:error}, we consider the case of a bandpass mismatch either due to just the sampling resolution (with $\tau_s$ resolution = 1 GHz) or due to measurement uncertainties (with $\sigma = 0.0089$ and sampling resolution $\tau_s=0.5$ GHz). In the latter case, together with the realization of CMB and noise, we also vary the realization for the uncertainties of the bandpasses measurements. To reduce the computing time to generate these maps, we halved the sampling rate of the observations (from the \textit{LiteBIRD} default, i.e., 19 Hz, to 9.5 Hz). We have checked that this approximation results in a effect on the maps smaller than the one caused by the systematic under study (the impact of changing the sampling rate is about 5\% of the systematic residual at map level).

We thus derive the NILC weights for  $m_{\rm{out}}$/$m_{\rm{templ}}$ associated to each channel $j$: $w^{\mathrm{NILC}}_{j, \rm{out/templ}}$. The residual power spectrum is computed as:

\begin{equation}\label{eq:clres_ilc}
\begin{split}    
C^{BB,\rm{res}, \mathrm{NILC}}_{\ell} = C^{BB}\left[\sum_j (m^j_{\rm{out}} \cdot w^{\mathrm{NILC}}_{j, \rm{out}} - m^j_{\rm{templ}} \cdot w^{\mathrm{NILC}}_{j, \rm{templ}} ) \cdot m_{\rm{mask}} \right]/ f_{\rm{sky}}(B^{\rm{max}}_{\ell} p_{\ell})^2,
\end{split}
\end{equation}
where $m^j_{\rm{out/templ}} \cdot w^{\mathrm{NILC}}_{j, \rm{out/templ}}$ is the CMB + foreground (+systematics) residual from component separation, without noise. 

As done before, all maps have been brought to the resolution of the largest FWHM, $B^{\rm{max}}_{\ell}$. 
We use the same likelihood as in Eq.~\ref{eq:L}, $\mathcal{L}(\tilde{C}^{BB,  \mathrm{NILC}}_{\ell}|C^{BB}_{\ell}(r)+ C_{\ell}^{BB, \rm{noise}})$, where:
\begin{equation}\label{eq:clBB_ilc}  
\tilde{C}^{BB, \mathrm{NILC}}_{\ell} = C^{BB, \rm{fid}}_{\ell} + C^{BB, \rm{res}, \mathrm{NILC}}_{\ell} + C^{BB, \rm{noise}}_{\ell},
\end{equation}
and $C^{BB, \rm{noise}}_{\ell}$ is the power spectrum of the NILC noise residuals when applied to the dataset with systematics $m^{B}_{\rm{out}}$.

The $\ell$ range used in this case is $2<\ell<150$, restricting the range to less than ${3 \times \texttt{NSIDE} - 1}$.\footnote{This is slightly different from what has been done in previous sections but we have verified that earlier results do not change significantly when evaluating the likelihood between $2<\ell<150$.} This choice speeds up the component separation step without losing constraining power on primordial tensor modes in the $B$-mode power spectrum.

The overall degradation of the sensitivity on the tensor-to-scalar ratio $\Delta  \hat{r}$ is computed from the distribution of $\Delta r_i$ for the different $50$ realizations as~\cite{Carralot:2024eau}:

\begin{equation}
    \Delta \hat{r}  = \sqrt{\mu_{\Delta r_i}^2 + \text{Var}[\Delta r_i]} = \sqrt{\langle \Delta r_i^2 \rangle}.
    \label{eq:delta_r_tot}
\end{equation}
where $\mu_{\Delta r_i}$ is the average over $\Delta r_i$.

We obtain:
\begin{itemize}
    \item for the case with sampling resolution 1 GHz: $\Delta \hat{r} = 1.1 \times 10^{-9}$, their standard deviation is $\text{std}(\Delta r_i) = 2.2 \times 10^{-10}$;
    \item for the case with $\sigma = 0.0089$ and sampling resolution 0.5 GHz:  $\Delta \hat{r} = 3.0 \times 10^{-8}$, their standard deviation is $\text{std}(\Delta r_i) = 2.5 \times 10^{-8}$. 
\end{itemize}

Such results demonstrate that derived requirements on sampling and bandpass uncertainties fully match the \textit{LiteBIRD} $\Delta r$ budget allocated per systematic effect when performing a blind component separation analysis. 

Since the NILC methodology has been proven to be effective in removing the systematic residual, we proceed to check whether the requirement on $\sigma$ can be relaxed.\footnote{From the results of Section~\ref{sec:resolution} we expect also a resolution of 1.5 GHz (with no measurement errors) to be acceptable. We do not repeat the analysis also for this case, since a resolution of 1 GHz would anyway be more widely assumed for most studies and it should not be challenging to obtain in FTS measurements.} We simulate other 50 CMB + noise + bandpass noise realization with larger uncertainties on $\tau_s$ ($\sigma = 0.05$) and sampling resolution kept fixed to 0.5 GHz. Even in this case, after applying NILC and following the procedure described in this section, we find $\Delta \hat{r} = 1.2 \times 10^{-6}$ (std($\Delta r_i) = 1.0 \times 10^{-6}$). This indicates that the requirement on $\sigma$ can be relaxed to $\sigma \lesssim 0.05$ when using NILC. Such an outcome is aligned to findings of other similar studies \cite{Carralot:2024eau} as non-parametric component separation methods demonstrated to be more robust with respect to distortions of the foreground components. In our case, despite the distortion of the CMB signal being present in our set of simulations due to the bandpass uncertainty, it is subdominant with respect to the foreground distortion in $B$ modes.

\section{Conclusions}
In this work we study the effect of bandpass uncertainties (resolution and measurement error) on the estimate of $r$, in the context of the \textit{LiteBIRD} satellite and in the presence of a non-ideal HWP. We derive requirements on the level of bandpass resolution and measurement error such that the bias on $r$ does not exceed the threshold of 1\% of the budget allocated to systematic errors, i.e.\ $\Delta r < 6.5 \times 10^{-6}$~\cite{litebird_ptep, Mousset_prep}. These requirements are derived using only three reference \textit{LiteBIRD} frequency channels, due to the computational cost of our procedure, and for three bandpass shapes (top-hat, Chebyshev order 5 and Chevyshev order 3). We consider a foreground sky with non-spatially varying dust and synchrotron (the \texttt{d0s0 PySM} models). The impact of more complicated models is deferred to future work. We find that resolutions $\lesssim 1.5$ GHz and a Gaussian errors with $\sigma \lesssim 0.0089$ (for a resolution of 0.5 GHz) produce acceptable levels of map residuals and bias on $r$ for all three channels and bandpass shapes.

In a second part of the analysis, we restrict ourselves to the bandpass shape producing the most conservative results (Chebyshev order 3), and check that the derived requirements are still viable when systematic effects are injected in all the \textit{LiteBIRD} frequency channels. We compute the residual maps from all channels corresponding to a sampling resolution of 1 GHz and to a measurement error with $\sigma = 0.0089$ (and resolution 0.5 GHz). In both cases, we derive the total residual power spectrum and the corresponding bias on $r$:  $\Delta r = 6.5 \times 10^{-7}$ in the case with resolution 1 GHz (with no error) and $\Delta r = 4.7 \times 10^{-6}$ in the case with Gaussian error. Both cases are still below the threshold $\Delta r < 6.5 \times 10^{-6}$.

We have also verified that the requirements are still valid when performing a blind component separation (NILC) analysis. We have generated 50 different CMB, noise and bandpass noise realizations for the two requirements and performed NILC component separation. We find that the component separation step is more effective than our simple deprojection procedure, leading to $10^{-10} < \Delta r < 10^{-7}$. Since NILC is so effective in cleaning the systematic residual, we also did a test on 50 realizations with $\tau_s$ resolution = 0.5 GHz and Gaussian error with $\sigma = 0.05$. Using NILC, we get $\Delta \hat{r} = \sqrt{\mu_{\Delta r_i}^2 + \text{Var}[\Delta r_i]} = 1.2 \times 10^{-6}$. This suggests that the requirement on $\sigma$ can be relaxed to $\sigma \lesssim 0.05$ when applying NILC as a component separation method, due to its robustness against systematic effects that mainly distort the foreground components in the $B$-mode field.

\acknowledgments
We acknowledge the Hawk high-performance computing cluster at the Advanced Research Computing at Cardiff (ARCCA). This work is supported in Japan by ISAS/JAXA for Pre-Phase A2 studies, by the acceleration program of JAXA research and development directorate, by the World Premier International Research Center Initiative (WPI) of MEXT, by the JSPS Core-to-Core Program of A. Advanced Research Networks, and by JSPS KAKENHI Grant Numbers JP15H05891, JP17H01115, and JP17H01125.
The Canadian contribution is supported by the Canadian Space Agency.
The French \textit{LiteBIRD} phase A contribution is supported by the Centre National d’Etudes Spatiale (CNES), by the Centre National de la Recherche Scientifique (CNRS), and by the Commissariat à l’Energie Atomique (CEA).
The German participation in \textit{LiteBIRD} is supported in part by the Excellence Cluster ORIGINS, which is funded by the Deutsche Forschungsgemeinschaft (DFG, German Research Foundation) under Germany’s Excellence Strategy (Grant No. EXC-2094 - 390783311).
The Italian \textit{LiteBIRD} phase A contribution is supported by the Italian Space Agency (ASI Grants No. 2020-9-HH.0 and 2016-24-H.1-2018), the National Institute for Nuclear Physics (INFN) and the National Institute for Astrophysics (INAF).
Norwegian participation in \textit{LiteBIRD} is supported by the Research Council of Norway (Grant No. 263011) and has received funding from the European Research Council (ERC) under the Horizon 2020 Research and Innovation Programme (Grant agreement No. 772253 and 819478).
The Spanish \textit{LiteBIRD} phase A contribution is supported by MCIN/AEI/10.13039/501100011033, project refs. PID2019-110610RB-C21, PID2020-120514GB-I00, PID2022-139223OB-C21, PID2023-150398NB-I00 (funded also by European Union NextGenerationEU/PRTR), and by MCIN/CDTI ICTP20210008 (funded also by EU FEDER funds).
Funds that support contributions from Sweden come from the Swedish National Space Agency (SNSA/Rymdstyrelsen) and the Swedish Research Council (Reg. no. 2019-03959).
The UK  \textit{LiteBIRD} contribution is supported by the UK Space Agency under grant reference ST/Y006003/1 - "LiteBIRD UK: A major UK contribution to the LiteBIRD mission - Phase1 (March 25)."
The US contribution is supported by NASA grant no. 80NSSC18K0132.
We also acknowledge support from  the JSPS KAKENHI Grant Number 22K14054 and 24K21545, the INFN InDark initiative and the COSMOS network through the ASI (Italian Space Agency) Grants 2016-24-H.0.

\bibliographystyle{JHEP}
\bibliography{draft,Planck_bib}

\end{document}